\documentclass[times,twocolumn,preprint]{aastex631}

\usepackage{amsmath}
\usepackage{cases}
\usepackage{graphicx}
\usepackage{subfigure}
\usepackage{natbib}
\usepackage{color}
\usepackage{tabularx}
\usepackage{bm}
\usepackage{threeparttable}
\usepackage{scrextend}

\begin{document}

\title{{\large Microlensing Events in Five Years of Photometry from the Zwicky Transient Facility}}
\author[0009-0004-1650-3494]{Ruocheng Zhai}
\affiliation{Department of Astronomy, California Institute of Technology, 1200 E. California Blvd, Pasadena, CA 91125, USA}
\affiliation{Department of Astronomy, Tsinghua University, Beijing 100084, China}
\email{ruocheng.zhai@gmail.com}

\author[0000-0003-4189-9668]{Antonio C. Rodriguez}
\affiliation{Department of Astronomy, California Institute of Technology, 1200 E. California Blvd, Pasadena, CA 91125, USA}

\author[0000-0001-8317-2788]{Shude Mao}
\affiliation{Department of Astronomy, Tsinghua University, Beijing 100084, China}

\author[0000-0002-6406-1924]{Casey Y. Lam}
\affiliation{Observatories of the Carnegie Institution for Science, 813 Santa Barbara Street, Pasadena, CA 91101, USA}

\author[0000-0001-8018-5348]{Eric C. Bellm}
\affiliation{DIRAC Institute, Department of Astronomy, University of Washington, 3910 15th Avenue NE, Seattle, WA 98195, USA}

\author[0000-0003-1227-3738]{Josiah Purdum}
\affiliation{Caltech Optical Observatories, California Institute of Technology, 1200 E. California Blvd, Pasadena, CA 91125, USA}

\author[0000-0002-8532-9395]{Frank J. Masci}
\affiliation{IPAC, California Institute of Technology, 1200 E. California Blvd, Pasadena, CA 91125, USA}

\author[0000-0002-9998-6732]{Avery Wold}
\affiliation{IPAC, California Institute of Technology, 1200 E. California Blvd, Pasadena, CA 91125, USA}


\begin{abstract}
Microlensing has a unique advantage for detecting dark objects in the Milky Way, such as free-floating planets, neutron stars, and stellar-mass black holes. Most microlensing surveys focus on the Galactic bulge, where higher stellar density leads to a higher event rate. However, microlensing events in the Galactic plane have closer lenses and longer timescales, which leads to a greater chance of measuring microlens parallax, providing an additional constraint on the mass of the lens. This work searches for microlensing events in Zwicky Transient Facility (ZTF) Data Release 17 from 2018--2023 in the Galactic plane region. We find 124 high-confidence microlensing events and 54 possible events, all available online\footref{online_resources}. Thus, with two years of additional ZTF data in DR17, we have more than doubled the number of microlensing events (60) found in the previous three-year DR5 search. In the event selection, we use the efficient \texttt{EventFinder} algorithm to detect microlensing signals, which could be used for large datasets such as future ZTF data releases or data from the Rubin Observatory Legacy Survey of Space and Time (LSST). Using detection efficiencies of ZTF fields obtained from catalog-level simulations, we calculate the mean Einstein timescale to be $\langle t_\mathrm{E}\rangle=51.7\pm3.3$ days, smaller than previous results of the Galactic plane but within 1.5$\sigma$. We also calculate optical depths and event rates, although some caution is needed due to the use of visual inspection when creating our final sample. Spectroscopy of three possible candidates confirms their microlensing nature.
\end{abstract}. 

\shorttitle{}
\shortauthors{Zhai et al.}

\section{Introduction}

Dark celestial objects, such as stellar remnants and exoplanets, are challenging to probe by typical astronomical methods involving observations of their electromagnetic radiation. However, their gravitational fields bend the light passing by them through gravitational lensing (e.g., \citealp{Einstein_GLens, Paczynski1986}). Objects of large mass, like galaxy clusters, can produce strong distortion on the images of background sources, producing arcs or distorted galaxies on the image in an effect called ``strong lensing". However, the observer cannot resolve stellar-mass object images of a background source due to the resolution limit. However, we can probe the properties of dark lenses in our Galaxy by observing light curves of sources \citep*{Paczynski1986, Paczynski1996, Mao1991}. 

A microlensing event is detectable when the source trajectory has an impact parameter of the order of the angular Einstein radius:
\begin{equation}
    \theta_\mathrm{E} = \sqrt{\kappa M \pi_\mathrm{rel}},
\end{equation}
where $\kappa = 8.14\ \mathrm{mas}\ M_\odot^{-1}$, $M$ is the mass of the lens, and $\pi_\mathrm{rel} = 1\ \mathrm{au}\ (1/D_L - 1/D_S)$ is the relative parallax between the lens and the source. The microlens parallax $\pi_\mathrm{E} = \pi_\mathrm{rel}/\theta_E$ is introduced to project the motion of the observer to the lens plane. Its direction takes that of the source proper motion so that the microlens parallax vector is:
\begin{equation}
    \bm{\pi}_\mathrm{E} = (\pi_{\mathrm{E,N}}, \pi_{\mathrm{E,E}}) = \frac{\pi_\mathrm{rel}}{\theta_\mathrm{E}}\frac{\bm{\mu}_\mathrm{rel}}{\mu_\mathrm{rel}},
\end{equation}
where N and E refer to the North and East components of $\bm{\pi}_\mathrm{E}$.By analyzing the source trajectory observed in the lens plane, $\pi_\mathrm{rel}$ can be constrained \citep{Gould2004}. If we can observe the same event from the Earth and a satellite, the different source trajectories they see can be used to measure $\pi_\mathrm{E}$ \citep{Gould1994, Zang2020}. For events only observed from the Earth, observing the microlens parallax is only possible for events with long Einstein timescales. \cite{Mroz2020} showed that the average Einstein timescales ($t_\mathrm{E} = \theta_\mathrm{E}/\mu_\mathrm{rel}$) of Galactic plane events are longer ($t_E \sim 60$ days) than bulge ones ($t_E \sim 20$ days), suggesting that Galactic plane events are more likely to show detectable parallax effects.

However, observing a dense field is the most profitable since microlensing requires a very close alignment ($\sim$ mas) between the lens and the source. Thus, high cadence searches such as the Korea Microlensing Telescope Network (KMTNet, \citealt{KMT2016}) and Microlensing Observations in Astrophysics (MOA; \citealt{Sumi2016}) mainly focus on the Galactic bulge. The Optical Gravitational Lensing Experiment (OGLE; \citealt{Udalski2015}), however, covers a wide range of sky, observing the Galactic bulge, part of the Galactic plane, and the Small/Large Magellanic Clouds. It showed that the microlensing event rate towards the Galactic bulge is of the order $ 10^{-5}\ \mathrm{year^{-1}}$ \citep{Mroz2019}, while only $10^{-6}\ \mathrm{year^{-1}}$ towards the Galactic plane \citep{Mroz2020}.  \cite{Rodriguez2022} conducted a systematic search on the archive of the Zwicky Transient Facility (ZTF) Data Release 5 (DR5) focusing on the Northern Galactic plane (here, we define "Northern" to be within the grasp of ZTF, i.e., Decl.$>-28^\circ$, finding a total of 60 microlensing events, with a mean Einstein timescale $\langle t_\mathrm{E}\rangle = 61.0 \pm 8.3$ days. \cite{2023medford} independently searched for microlensing events in ZTF DR5 and found 60 events. The mean timescale of \cite{Rodriguez2022} is consistent with the result of \cite{Mroz2020} from events in the Southern Galactic plane. The ZTF microlensing sample in the Northern Galactic plane is an important supplement to OGLE-IV \citep{Mroz2020}, which conducted the only microlensing survey of the Galactic plane from the southern hemisphere.

In this paper, we search for microlensing events in the 5-year-long archive of ZTF photometry to build a larger sample for systematical analysis. The previous ZTF DR5 covered approximately 3 years of photometric data from 2018 to 2021, whereas, in this study, we use ZTF DR17, which covers approximately 5 years of data through 2023. It enables us to analyze newly transpired microlensing events and further study ones that could have begun brightening during the baseline of DR5 but did not return to quiescence until after the end of DR5.

The outline of the paper is as follows. Section \ref{sec:2} describes the ZTF data archive. In section \ref{sec:3}, we introduce our methodology to select and analyze microlensing events. In section \ref{sec:4}, we perform a simulation to calculate detection efficiencies and the mean Einstein timescale. We discuss the microlensing optical depth and event rate measured from our sample. We draw our conclusion in section \ref{sec:5}.

\section{Data}\label{sec:2}

ZTF is a time-domain survey using the Samuel Oschin 48-inch telescope with a 47 $\mathrm{deg^2}$ field-of-view camera at Palomar Observatory \citep{Graham2019, Masci2019, Bellm2019b, Dekany2020}. ZTF carries out a public survey in the entire northern sky in $g$ and $r$ bands \citep{Bellm2019a, Kupfer2021} with a 3-day cadence in Phase I (first year) \citep{Bellm2019b} and a 2-day cadence in Phase II. Some ZTF partnership data are not currently publicly available but will be published in the near future. However, the vast majority of private data is related to alerts and transient detections, and their inclusion does not strongly impact the sample of events in this paper or alter model parameters. The pixel size of the ZTF camera is $1^{\prime\prime}$, and the FWHM of the average delivered image quality is $2^{\prime\prime}.00$.

ZTF Data Release 17 contains public survey data up to 9 March 2023 and private data prior to 6 November 2021, both starting from 17 March 2018. We use all available (both public and private) data in the Galactic plane, accessing them using the Kowalski\footnote{\url{https://github.com/dmitryduev/kowalski}} datastore \citep{Duev2019} query tool through the Python package \texttt{penquins}\footnote{\url{https://github.com/dmitryduev/penquins}} developed for the ZTF collaboration, which uses the pipeline described in Section 3.2 of \cite{Bellm2019a} to produce PSF (point-spread function) fit catalogs. Light curves have a photometric precision ranging from 0.01 mag at 14 mag and up to 0.1-0.2 mag at 20-21 mag for the faintest objects. Because of the Galactic reddening, we start with analyzing 565,054,933 light curves with at least 30 points in 196 fields of the $r$-band data, typically with a greater signal-to-noise ratio than other bands ($g$ and $i$ band). 

\section{Methods}\label{sec:3}

The point-source point-lens (PSPL) model describes the simplest microlensing events involving a single point source and a single point lens. The magnified flux at time $t$ is given by:
\begin{equation}\label{eq:flux}
    F(t) = 
    \begin{cases}
    A(t; t_0, u_0, t_\mathrm{E})f_s + f_b \qquad & \mathrm{no parallax} \\
    A(t; t_0, u_0, t_\mathrm{E}, \pi_\mathrm{E,N}, \pi_\mathrm{E,E})f_s + f_b \qquad & \mathrm{parallax},
    \end{cases}
\end{equation}
in which $t_0$ is the time when the source is the closest to the lens, $u_0$ is the closest distance normalized to the Einstein radius $\theta_E$, the Einstein timescale $t_\mathrm{E}$ is the Einstein radius crossing time, $f_s$ is the source flux, and $f_b$ is the blending flux due to light from the lens and/or unrelated sources of light within the seeing disk. With a $(t_0, u_0, t_E)$ set, $f_s$ and $f_b$ can be deduced from a linear regression based on the light curve. Thus, we must search for the best parameters in the 3-dimensional space of $(t_0, u_0, t_E)$. However, considering the large amount of data in ZTF DR17, we must first search for possible microlensing signals in light curves using the PSPL model and obtain reasonable model parameters. Here, we use \texttt{MulensModel}\footnote{\url{https://rpoleski.github.io/MulensModel/index.html}} \citep{MulensModel} to calculate the magnification and fit the PSPL model. To fit a light curve with the PSPL model, we minimize the $\chi^2$:
\begin{equation}\label{eq:chi2}
    \chi^2 = \sum_{i = 1}^{N}\left(\frac{F_i - F_{i,\mathrm{model}}}{\sigma_i}\right)^2,
\end{equation}
where $F_i$ is the observed flux, $F_{i,\mathrm{model}}$ is the flux calculated by the model (eq. \ref{eq:flux}), and $\sigma_i$ is the error bar. Specific optimization algorithms will be introduced in the following subsections. For those light curves having gone through the above cuts, we visually inspect them in multi-band data to confirm whether they are microlensing events.

\subsection{Skewness-Von Neumann Space}

Following the previous work of \cite{Rodriguez2022}, we first calculate the skewness $\gamma$ and Von Neumann statistic $\eta$ of all $r$-band light curves in DR17. They are defined as:

\begin{equation}
    \begin{aligned}
        \gamma &= \sum^N_{i=1}\frac{(x_i-\bar{x})^3}{N\hat{\sigma}^3}, \\
        \eta &= \sum^N_{i=2}\frac{(x_i-x_{i-1})^2}{(N-1)\hat{\sigma}^2},
    \end{aligned}
\end{equation}
where $x_i$ is a data sequence, i.e., the magnitude in our case, and $\hat{\sigma}$ is the standard deviation of $x_i$. The skewness parameter $\gamma$ becomes negative if a brightening signal appears in a light curve (magnitude becomes smaller). The Von Neumann statistic $\eta$ symbolizes the significance of long-term deviations from the baseline.

Based on the microlensing events confirmed by \cite{Rodriguez2022}, we calculate the $\gamma$ and $\eta$ parameters of their DR17 light curves to develop our cuts in the skewness-Von Neumann space:
\begin{equation}
    \gamma < 0;\ \log_{10}(-\gamma) > 3\log_{10}\eta + 0.3.
\end{equation}

\begin{table*}
    \renewcommand\arraystretch{1.25}
    \centering
    \caption{Cuts on ZTF DR17 light curves. Our automatic selection methodology includes three steps: 1. Skewness-Von Neumann Space; 2. EventFinder algorithm; 3. PSPL modeling. It reduces the number of light curves needed to inspect manually to 3,201 (a factor of  ~200,000 smaller than the original dataset). Of the 124 high confidence and 54 possible events, 51 are discovered in the ZTF DR5 search \citep{Rodriguez2022}.}
    \begin{tabular}{l c r}
    \hline
    \hline
    Criteria & Remarks & Number Left \\
    \hline
    All light curves in ZTF DR17 & Light curves with more than 30 points. & 565,054,933 \\
    \hline
    \textbf{Step 1:} skewness-Von Neumann Space & & \\
    $\gamma < 0;\ \log_{10}(-\gamma) > 3\log_{10}\eta + 0.3$ & Light curves with significant brightening signals. & 805,088 \\
    \hline
    \textbf{Step 2:} EventFinder algorithm & & \\
    $\Delta \chi^2 > 200$ & The EventFinder microlensing model fits well in the EventFinder window. & \\
    $\chi^2_\mathrm{out} < 4$ & Light curves are flat outside the EventFinder window. & 328,086 \\
    \hline
    Model converges & \texttt{scipy} optimisation of the PSPL model returns a successful status. & 136,939 \\
    \hline
    \textbf{Step 3:} PSPL modelling & & \\
    $N_\mathrm{pts,lens} \geq 6$ & Enough points within the lensing window. & \\
    amplitude $\geq$ 0.2mag & Remove small-rise false positives. & \\
    $\chi^2_\mathrm{red, lens} < 4$ & PSPL fits well in the lensing window in terms of reduced $\chi^2$. & \\
    $\chi^2_\mathrm{red, whole} < 4$ & PSPL fits well for the whole light curve in terms of reduced $\chi^2$. & \\
    $t_0 \pm 0.3 t_\mathrm{E}$ fully covered in data & Remove events with baseline outside DR17 data. & 3,201\\
    \hline
    Visual inspection & Choose light curves with microlensing signatures. & 321\\
    Multi-band data analysis & Fit with multi-band data in DR17. & \\
    High-confidence events & Objects that are confirmed as microlensing events. & 124 \\
    Possible events & Objects that have microlensing features but cannot be confirmed. & 54 \\
    \hline
    \end{tabular}
    \label{cuts}
\end{table*}

\begin{figure*}[htb]
    \centering
    \includegraphics[width=0.95\linewidth]{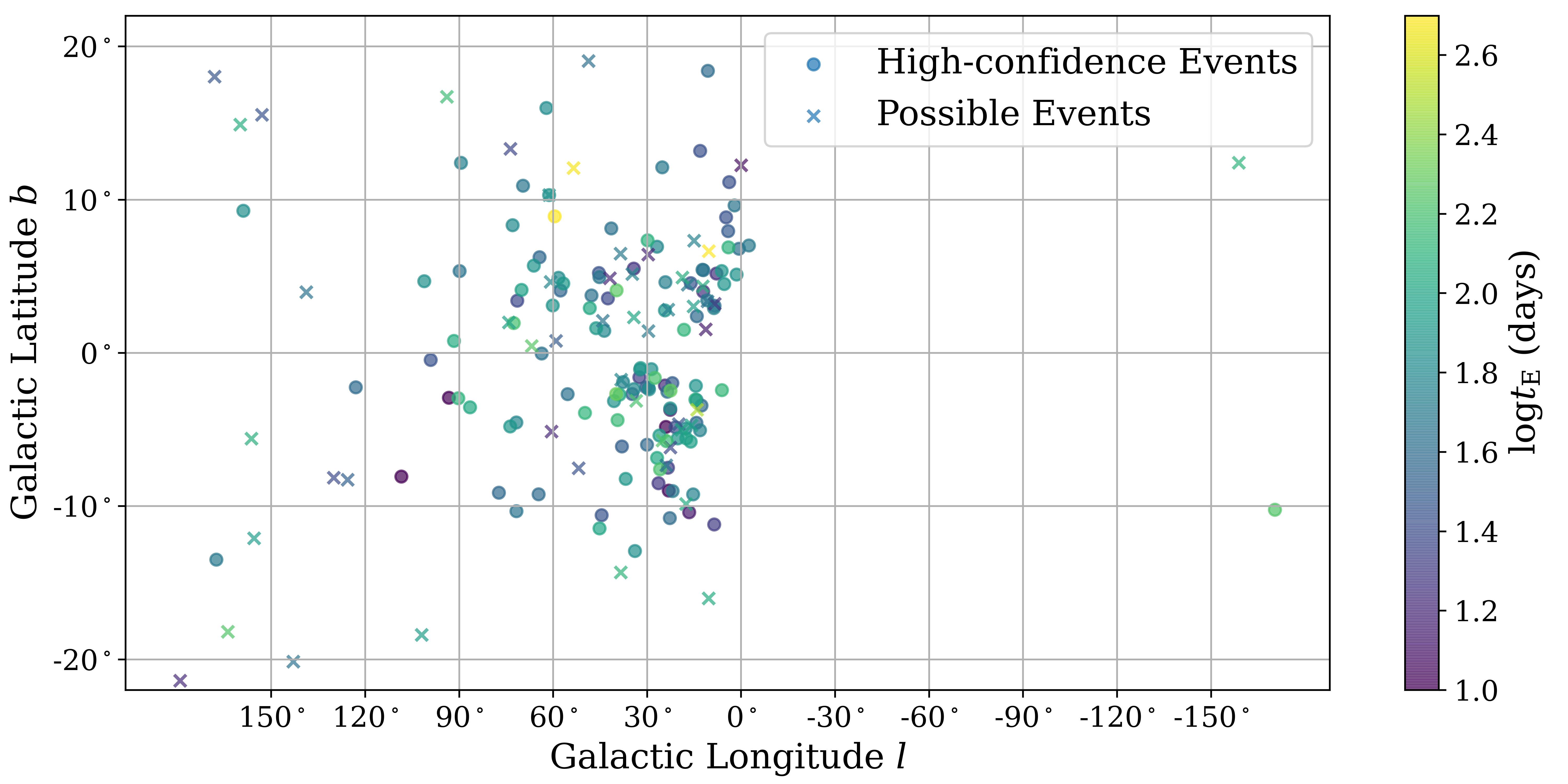}
    \caption{124 high-confidence and 54 possible microlensing events are plotted in Galactic coordinates as dots and crosses, respectively. Most high-confidence events are located at low Galactic latitudes and near the Galactic bulge. However, some events are located in the anti-bulge direction. In contrast, a greater proportion of the possible events are located at higher Galactic longitudes, indicating some are not real. Different colors of dots represent the magnitude of the Einstein timescales $t_\mathrm{E}$ of the events.}
    \label{gal}
\end{figure*}

\subsection{EventFinder Algorithm}

\cite{EventFinder} developed an EventFinder algorithm, simplifying the magnified flux to only determined by $(t_0,t_\mathrm{eff} \equiv u_0 t_\mathrm{E})$, while $u_0$ is constrained to $u_0 \to 0$  (high-magnification) or $u_0 = 1$, corresponding to $j = 1,\ 2$, respectively. $t_0,t_\mathrm{eff}$ are set to fixed values to produce approximations of light curves:
\begin{equation}
    F(t) = f_1A_j[Q(t; t_0, t_\mathrm{eff}] + f_0,\ j = 1,\ 2,
\end{equation}
\begin{equation}
    Q(t; t_0, t_\mathrm{eff}) \equiv 1+\left( \frac{t-t_0}{t_\mathrm{eff}} \right)^2,
\end{equation}
where:
\begin{equation}
    \begin{aligned}
        A_{j=1} &= Q^{-1/2},& &\mathrm{for}\ u_0 \to 0, \\
        A_{j=2} &=\frac{Q+2}{Q(Q+4)},& &\mathrm{for}\ u_0 = 1.
    \end{aligned}
\end{equation}
When $u_0 \to 0$, $f_1 \to F_\mathrm{max}$ (the maximum flux of a light curve) and when $u_0 = 1$, $(f_1,f_0) = (f_s, f_b)$.

We adopt 2-D grids of $(t_0, t_\mathrm{eff})$ from \cite{EventFinder}, conducting a linear regression to decide the two flux parameters ($f_1,\ f_0$) to detect whether a light curve has a signal similar to microlensing events. Also, EventFinder helps to determine the initial guess for the PSPL model fitting. We define:

\begin{equation}
    \Delta \chi^2 = \left(\frac{\chi^2_\mathrm{flat}}{\chi^2_
    \mathrm{mulens}} - 1\right)N_\mathrm{window},
\end{equation}
in which we choose a window of $(t_0 - 5t_\mathrm{eff}, t_0 + 5t_\mathrm{eff})$, $\chi^2_\mathrm{flat}$ and $\chi^2_
\mathrm{mulens}$ are calculated with a flat model and an EventFinder microlensing model according to the Equation (\ref{eq:chi2}), respectively, and $N_\mathrm{window}$ is the number of data points within the window. We set a $\Delta\chi^2 > 200$ threshold by calculating DR5 confirmed events. 

To remove light curves of variable stars, we calculated the reduced chi-square outside the window:
\begin{equation}
    \chi^2_\mathrm{out} = \frac{1}{N_\mathrm{out}-1}\sum_{i = 1}^{N_\mathrm{out}}\left(\frac{F_i - \bar{F}}{\sigma_i}\right)^2,
\end{equation}
where the window is at least 360 days, i.e., if $10t_\mathrm{eff} < 360\ \mathrm{days}$, we use a 360-day window to calculate $\chi^2_\mathrm{out}$, and $N_\mathrm{out}$ is the number of data points outside the window. We set a threshold of $\chi^2_\mathrm{out} < 4$ by calculating the $\chi^2_\mathrm{out}$ of events from \cite{Rodriguez2022}.

\subsection{PSPL Model Fitting}

After selecting events by the EventFinder and estimating $(t_0, t_\mathrm{eff})$ of a light curve, we set $u_0$ calculated from the magnification derived from the maximum and minimum flux of it as the initial guess of the PSPL model fitting. We use the Nelder-Mead method \citep{Singer2009} to find the best-fit parameters.

We implement several cuts based on the PSPL best-fit model. First, we define a lensing window
of 360 days centered at $t_0$. We remove light curves with fewer than 6 points within the lensing window, as in \cite{Mroz2017}. Also, light curves with an amplitude less than 0.2mag are excluded since they are often false positives (e.g., Be star outbursts). To decide whether the PSPL model fits well, we consider reduced chi-squared within the window as well as of the whole light curve:
\begin{equation}
    \chi_\mathrm{red}^2 = \frac{1}{N-\mathrm{d.o.f}}\sum_{i = 1}^{N}\left(\frac{F_i - \bar{F}}{\sigma_i}\right)^2,
\end{equation}
where $N$ is the total number of data points of a light curve. With the confirmed events in DR5, we set a $\chi_\mathrm{red}^2 < 4$ threshold for both within the lensing window and the whole light curve. Lastly, we choose the light curves that have observations before and after the time range of $t_0 \pm 0.3 t_\mathrm{E}$, ensuring events that are rising but do not have a decline covered by DR17, or vice versa will not make the final sample.

\subsection{Multi-band Modelling}

After the PSPL model fitting cuts, we visually inspect 3,201 events having passed and select 321 events with significant microlensing features. Also, ZTF data exhibit some systematic false positives due to imaging artifacts or calibration errors. These false positives can be distinguished by their appearance in multiple light curves of nearby stars, and we manually remove them. We tried to acquire multi-band data for all of the remaining candidates. ZTF typically observes an object in $g$, $r$, and $i$-bands. However, for most objects, $i$-band observation only covers a short duration, and we abandon the $i$-band data of most candidates. Only those data covering microlensing signals well are used in the fitting. Because ZTF fields can overlap, we use data from all fields at the position of a given event for multi-band modeling in order to constrain parameters better. 

By introducing multi-band data into the fitting, we can remove those events with significant asymmetry and differences between $g$-band and $r$-band. These events are typically dwarf novae or supernovae. Also, we discard events with only half of the microlensing peak observed since they cannot be distinguished from false positives (e.g., dwarf nova outbursts, Be star outbursts, or supernovae), with only objects getting fainter after outbursts. We classified microlensing events into two classes: high-confidence and possible. The high-confidence events can be confirmed as microlensing events since ZTF covered their peak well with large signal-to-noise ratios. While possible events only sparsely cover the peak, suffer considerable noise, lack sufficient baseline data, and thus, cannot be firmly established.

We show the results of every step of cuts and the final microlensing events in table \ref{cuts}. Table \ref{parms_high} and \ref{parms_possible} show lists of high-confidence and possible events.

\subsection{MCMC Parameter Exploration}\label{sec:3.5}

To estimate the best values of model parameters, we use Markov Chain Monte Carlo (MCMC) to explore the parameter space and the posterior distribution of all parameters. With the PSPL parameters found by the multi-band analysis, we use the package \texttt{emcee} \citep{emcee} to conduct an MCMC analysis on all chosen events. Since the MCMC exploration may produce a negative blending flux, we impose a prior on $f_b$ \citep{Mroz2020}:
\begin{equation}
    \mathcal{L}_\mathrm{prior}=\left\{
    \begin{aligned}
        &1 \quad &f_b \geq 0, \\
        &\exp\left(-\frac{f_b^2}{2\sigma^2}\right) \quad &f_b < 0,
    \end{aligned}
    \right. 
\end{equation}
where $\sigma = f_\mathrm{min}/3$, and $f_\mathrm{min}$ is the flux corresponding to the magnitude $m = 21$. For the static model, we constrain $u_0$ to be positive considering symmetry. We set it free in parallax fitting because the parallax model may have degeneracy in positive and negative $u_0$ \citep{Gould2004}.

To select events whose parallax model fits significantly better than the static PSPL model, we empirically choose a threshold of $\Delta\chi^2 > 16$. Also, we check the cumulative $\chi^2$ and parameter corner plots of the selected parallax events to ensure the lensing window contributes to the lower $\chi^2$ and the existence of degenerate solutions. For events with degeneracy, we constrain the MCMC parameters based on the corner plot to obtain degenerate solutions separately.

Table \ref{parms_high} shows the MCMC results of all high-confidence events and parallax models of some events.

\section{Results and Discussion}\label{sec:4}

\subsection{Simulation}

We inject microlensing events into light curves in the field where events are detected to evaluate the ability of ZTF fields to detect microlensing events and the intrinsic distribution of the microlensing event rate. For a microlensing event that can be well-fitted by the PSPL model, we identify its field. We select 5,000 objects from each quadrant of the 16 CCDs for each field detected with microlensing events and generate 320,000 objects per field. 

For each object, we select the microlensing parameters from a flat distribution of $(t_0, u_0, \log t_\mathrm{E})$, in the range of $t_0\sim(t_\mathrm{min},t_\mathrm{max}),\ u_0\sim(0,1),\ \log t_\mathrm{E}\sim(0,3)$, where $t_\mathrm{min}$ and $t_\mathrm{max}$ are the time of the first and last epochs of an object and $t_\mathrm{E}$ is in days.

After generating an event, we apply the 3-step algorithm selection methodology in section \ref{sec:3}. In this way, we can decide whether our methods can detect a simulated event. We compared the detection efficiency curves of fields plotted in Figure 2 of \cite{Rodriguez2022}. We find the maximum value of detection efficiencies of each field is larger, and the detection efficiency curves drop slower on the long Einstein timescale side since the longer observation time ($\sim 1,786$ days) gives rise to better detection of the long Einstein timescale events.
\begin{figure}
    \centering
    \includegraphics[width = 1.0\columnwidth]{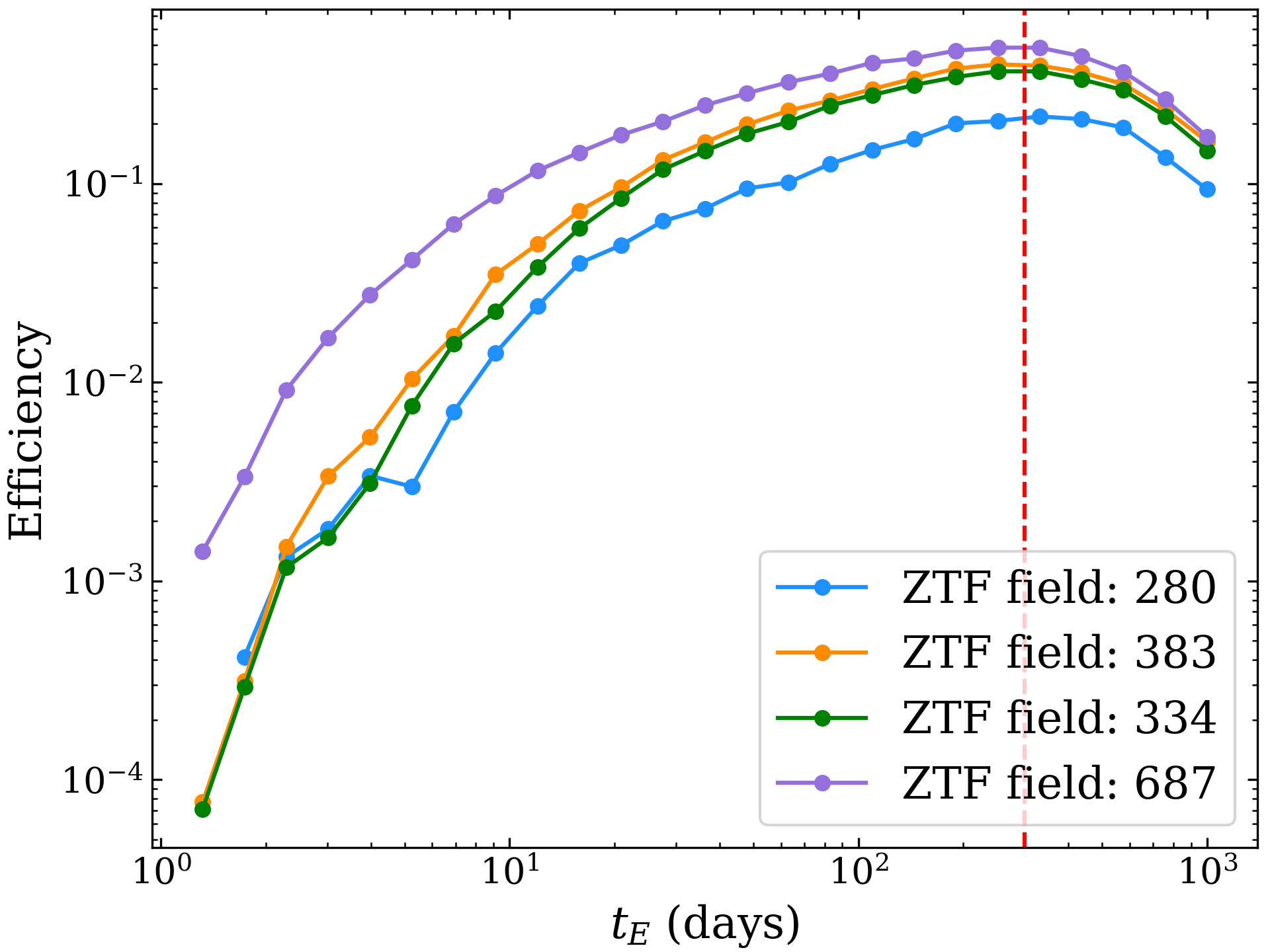}
    \caption{Detection efficiency curves for fields that were also shown in \protect\cite{Rodriguez2022}, updated by the DR17 objects with longer light curves. The dashed red line marks the $t_\mathrm{E}$ ($\sim 250$ days) where the detection efficiencies drop dramatically in DR5 due to the shorter baseline coverage.}
    \label{fig:field_det_eff}
\end{figure}

\subsection{Einstein Timescale Distribution}

To statistically analyze Einstein timescales, we first need to correct the weight of each event we have detected with the detection efficiencies. Each event represents $1/\epsilon(t_{\mathrm{E},i})$ events that actually happen, where $\epsilon(t_{\mathrm{E},i})$ is the detection efficiency of the $i\mathrm{th}$ event in its field. We only consider events of non-negligible detection efficiencies in the following analyses. By bootstrap sampling, we calculate the mean Einstein timescale $\langle t_\mathrm{E}\rangle$ {bootstrap sampling \citep{Efron1982}. We randomly sample from the detected Einstein timescales weighted by $1 / \epsilon(t_{\mathrm{E},i})$ repeatedly. For each Einstein timescale in a sample, we draw a value from the normal distribution based on its value and error and calculate the mean of the whole bootstrap sample. We obtain the distribution of the mean Einstein timescale from all the samples we generate. We remind the reader that it is fully possible to reproduce the detection efficiency curves presented here. We append a column of detection efficiencies associated with each event in Table \ref{parms_high} (see Appendix \ref{app:2}) and emphasize that the small amount of ZTF private data does not significantly affect the efficiency curves one can reproduce with the public data.

We find a mean Einstein timescale of $\langle t_\mathrm{E}\rangle = 51.7 \pm 3.3\ \mathrm{days}$, lower than the ZTF DR5 result (for the Northern Galactic plane) $\langle t_\mathrm{E}\rangle = 61.0 \pm 8.3\ \mathrm{days}$ \citep{Rodriguez2022}, as well as OGLE (for the Southern Galactic plane) $\langle t_\mathrm{E}\rangle = 61.5 \pm 5.0\ \mathrm{days}$ of \citep{Mroz2020}. However, our mean Einstein timescale is much longer than the Galactic bulge ($\sim 20$ days, \citealp{Mroz2019}). We also calculate the log-normal mean Einstein timescale $\langle t_\mathrm{E}\rangle_{\log}$ \citep{Wyrzykowski2015} which is less susceptible to outliers, giving $41.0 \pm 2.7\ \mathrm{days}$. We calculate the median value of Einstein timescale with the same method: $\tilde{t}_\mathrm{E} = 42.04  \pm 2.71$ days. The histogram in Figure \ref{fig:te_hist} shows the number of events normalized by the detection efficiencies vs. Einstein timescales.
\begin{figure}
    \centering
    \includegraphics[width = 1.0\columnwidth]{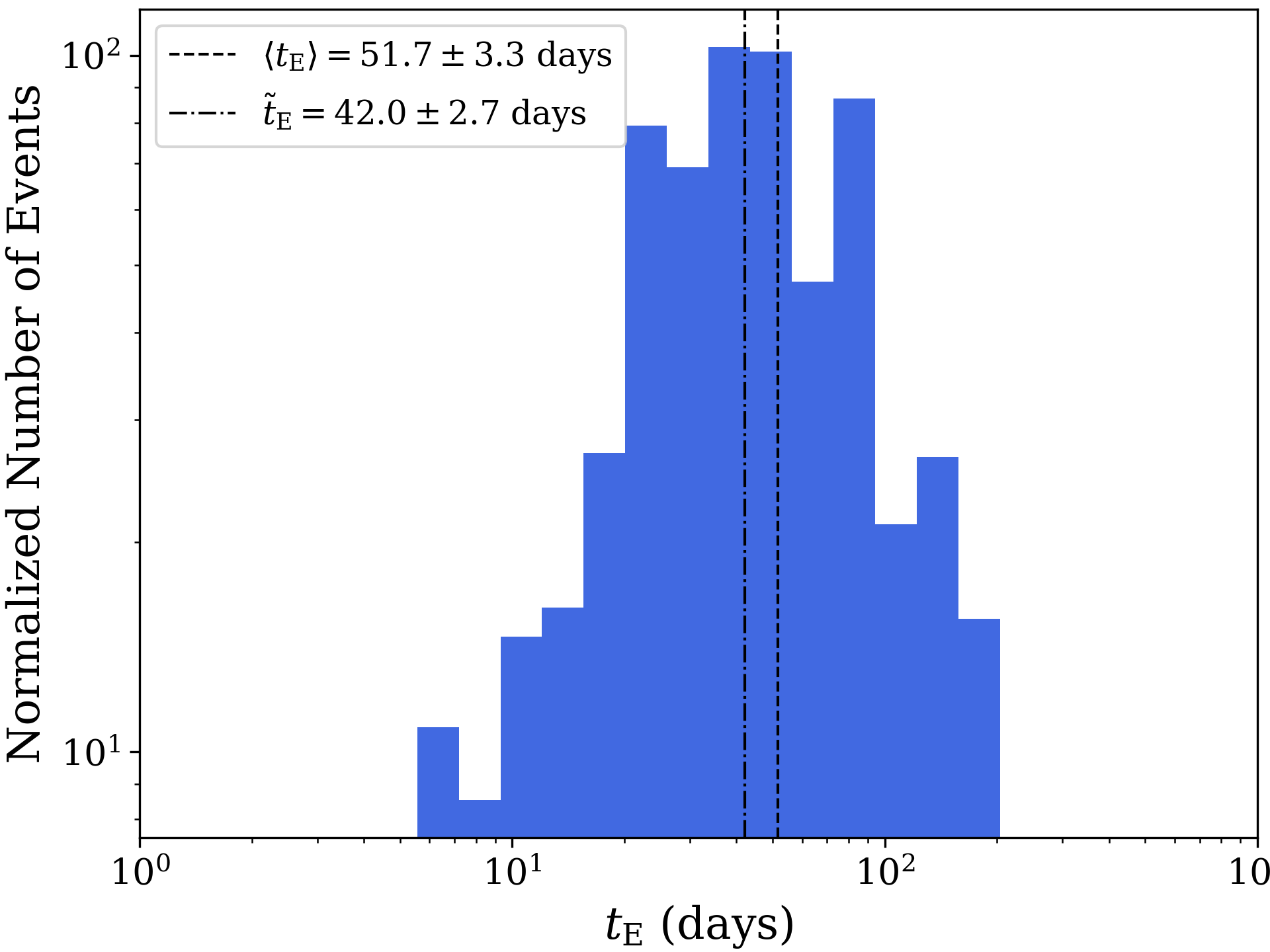}
    \caption{The histogram shows the event number normalized by the reciprocal of detection efficiencies with respect to Einstein timescales (i.e., the corrected histogram of $t_\textrm{E}$). The vertical dashed line marks the mean Einstein timescale, and the vertical dashed-dot line marks the median.}
    \label{fig:te_hist}
\end{figure} 

To evaluate the dependence of the mean Einstein timescales of different longitudes, we calculate mean Einstein timescales in five bins: $0^\circ < l < 15^\circ$, $15^\circ < l < 25^\circ$, $25^\circ < l < 40^\circ$, $40^\circ < l < 70^\circ$, $70^\circ < l < 180^\circ$, with approximately equal number in each bin. We use bootstrap sampling in each bin, weighting each event by the reciprocal of its detection efficiency. We show the result in the Figure \ref{fig:te_mean_l}. We find the mean Einstein timescales of events of low and high $l$ are smaller than those of medium $l$ events. When checking the high $l$ events, we note there are two events with $t_\mathrm{E} < 10\ \mathrm{days}$ in $70^\circ < l < 180^\circ$. If we exclude them, the mean Einstein timescale rises significantly, as shown by the orange dashed error bars in Figure \ref{fig:te_mean_l}. This indicates that a small sample of short Einstein timescale events can greatly influence the mean Einstein timescale since their small detection efficiencies give them large weights.
\begin{figure}
    \centering
    \includegraphics[width = 1.0\columnwidth]{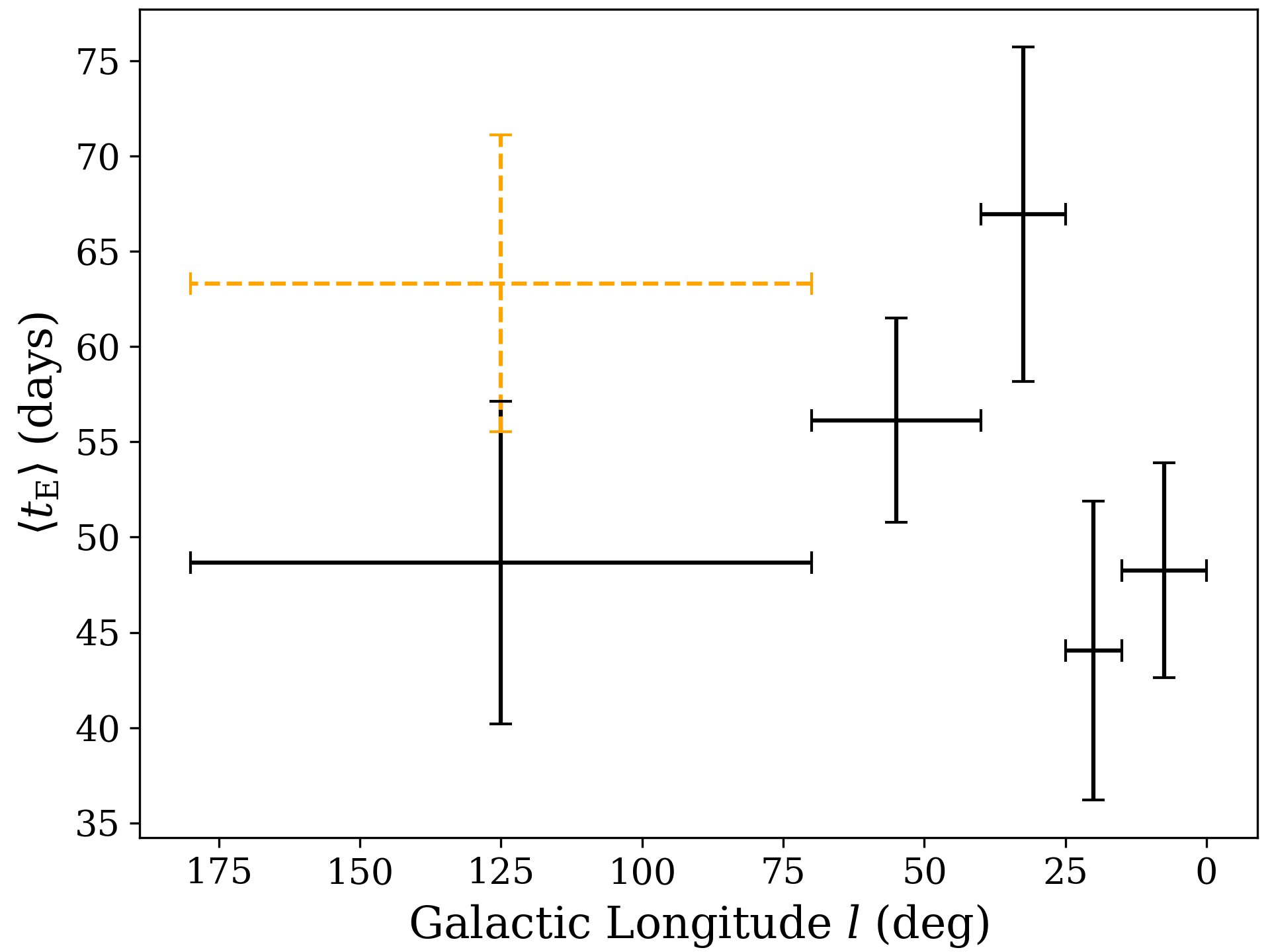}
    \caption{This figure shows the mean Einstein timescales calculated in different Galactic longitude ranges. Horizontal error bars represent the widths of bins, while vertical error bars are standard deviations of mean Einstein timescales. The orange dashed error bars are calculated without $t_\mathrm{E}<10\ \mathrm{days}$ events.}
    \label{fig:te_mean_l}
\end{figure} 
We plot the averaged detection efficiency curves of fields in each bin in Figure \ref{fig:field_det_eff_group} to check the performance of ZTF fields at different longitudes.
\begin{figure}
    \centering
    \includegraphics[width = 1.0\columnwidth]{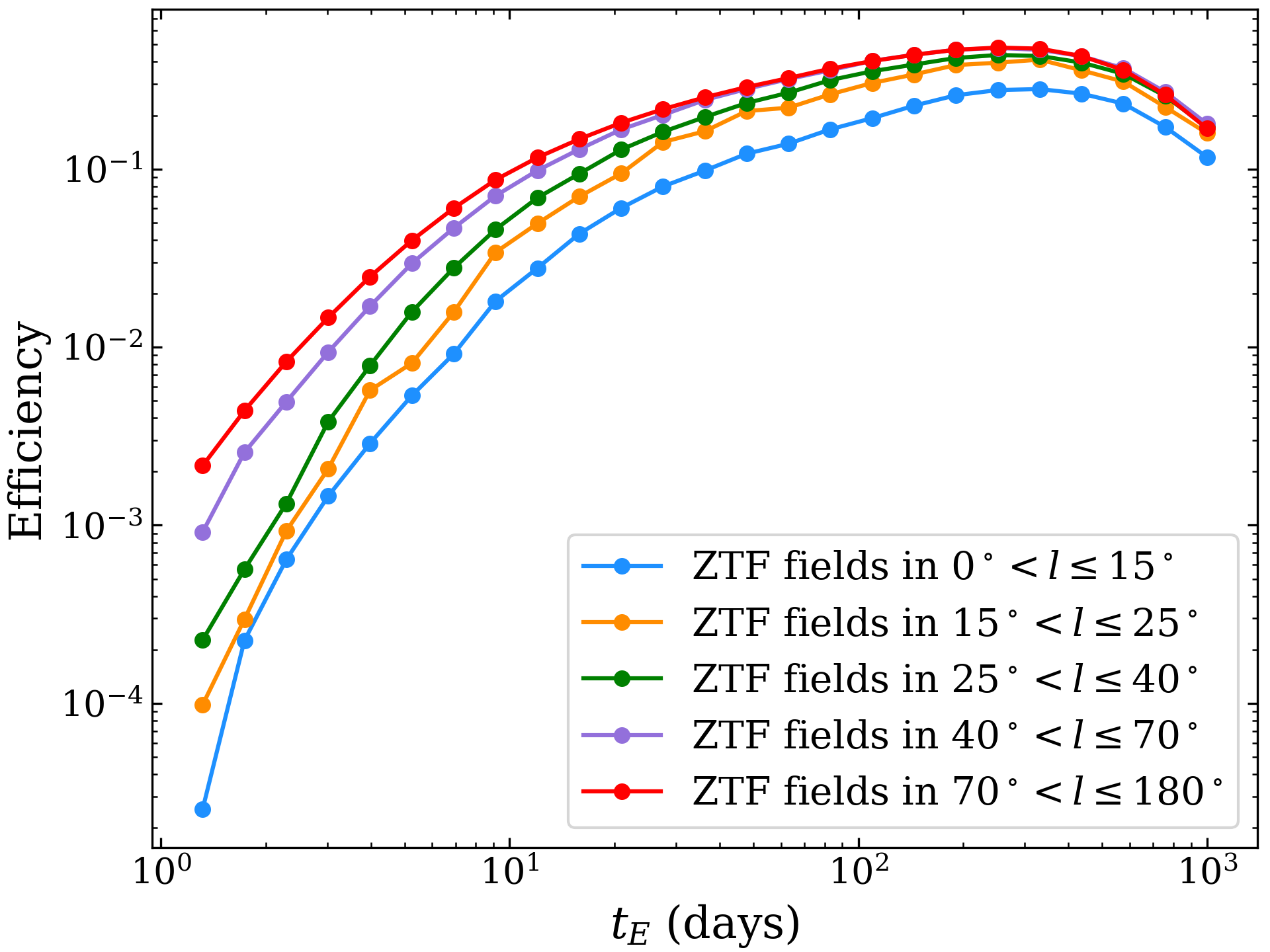}
    \caption{For each Galactic longitude bin, we average over the detection efficiencies of fields. Detection efficiencies rise as the Galactic longitude increases.}
    \label{fig:field_det_eff_group}
\end{figure}
As the longitude $l$ increases, the detection efficiencies also rise at the small Einstein timescale because of better coverage as the fields are up longer in the sky and fewer sources at large $l$. This explains why the two $t_\mathrm{E} < 10$ days events with non-negligible detection efficiencies are at high longitudes. Also, beyond the 3-step algorithm selection methodology, we find that when visually inspecting microlensing candidates, those with small $t_\mathrm{E}$ tend to have few observations around peaks of the light curves, making it hard to distinguish them from false positives. Hence, with the current cadence of ZTF, we may overestimate the mean Einstein timescales.

\subsection{Microlensing Optical Depth and Event Rate}

\begin{figure*}
    \centering
    \includegraphics[width = 1.0\columnwidth]{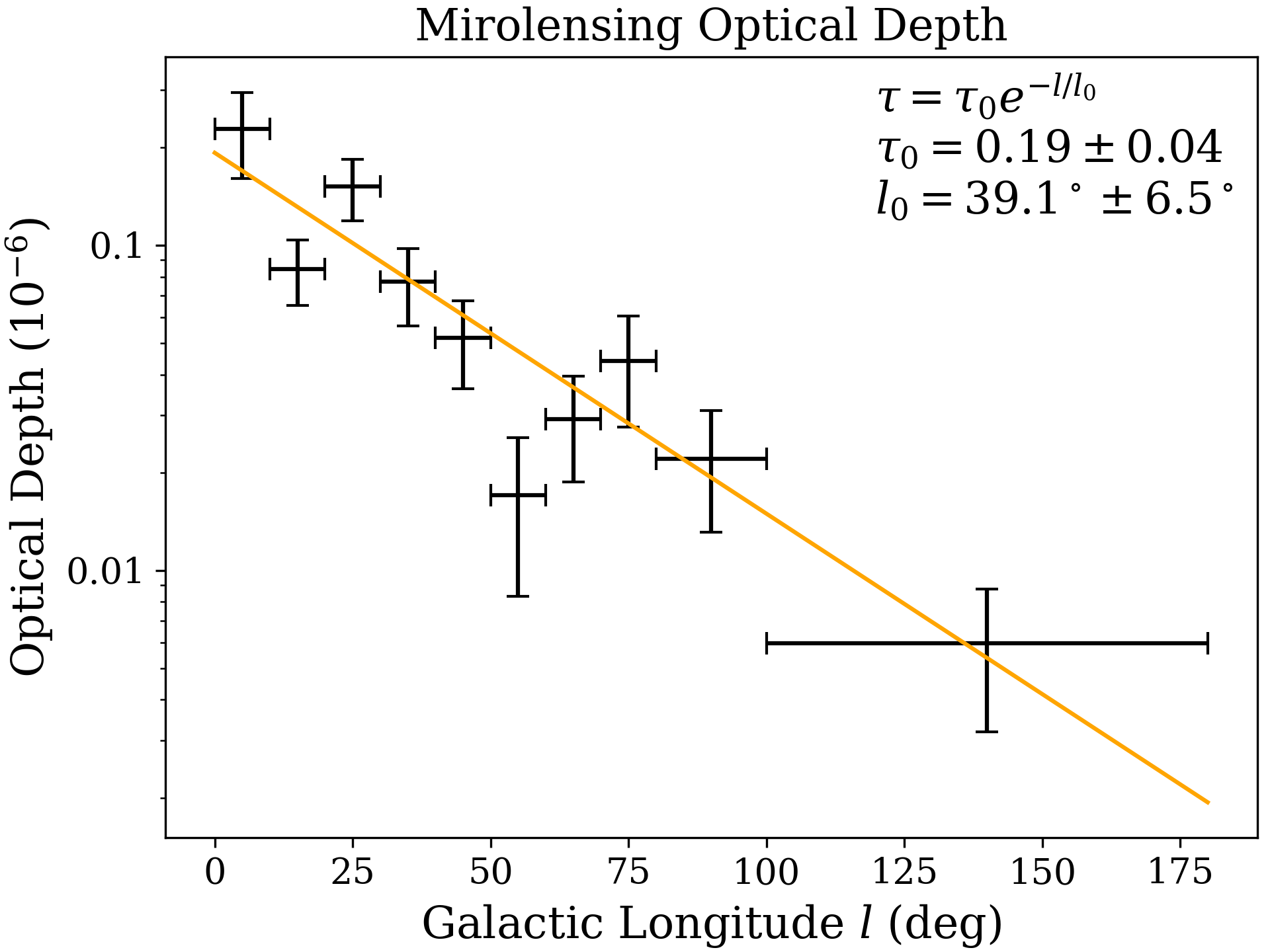}
    \includegraphics[width = 1.0\columnwidth]{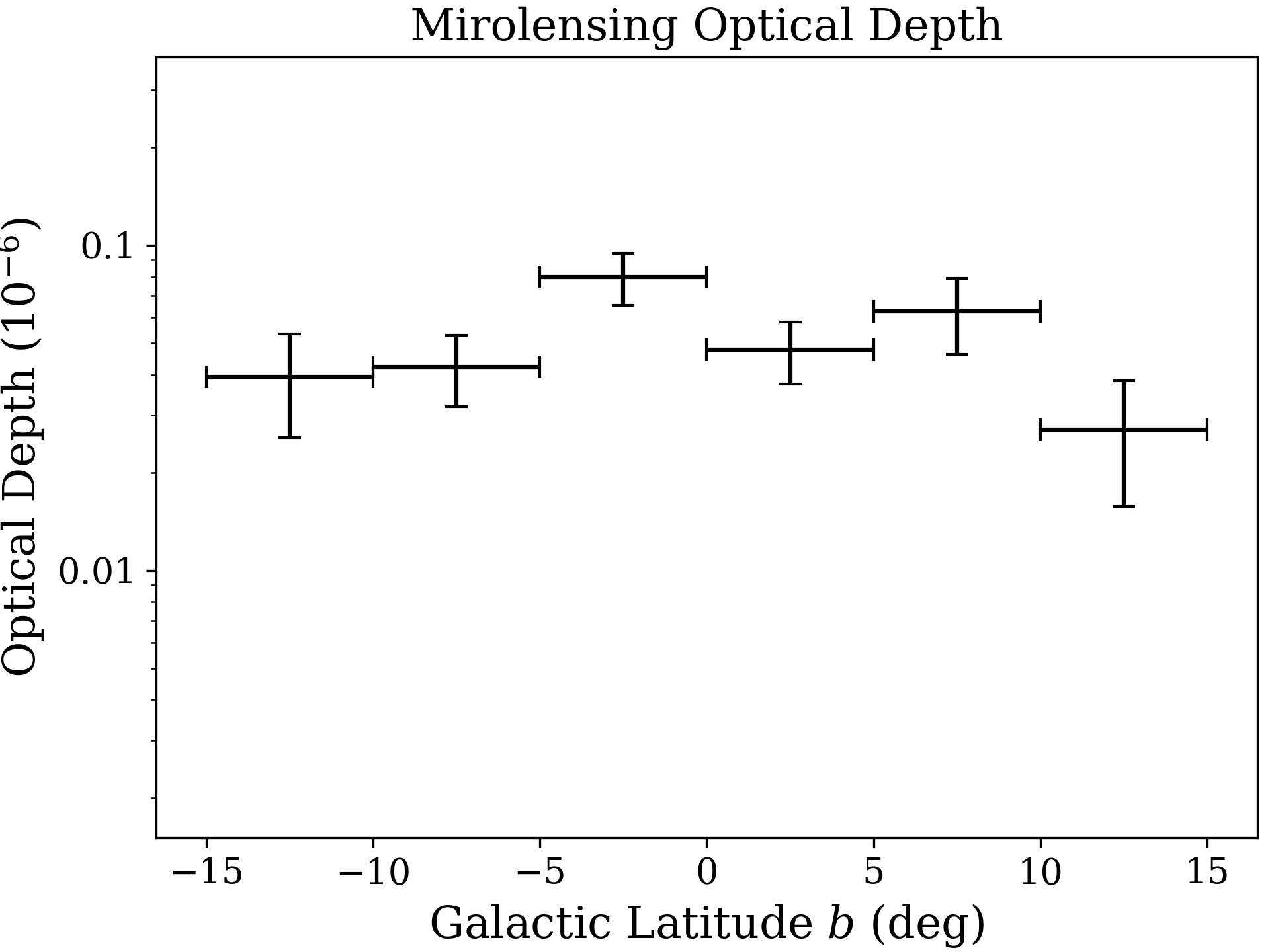}
    \includegraphics[width = 1.0\columnwidth]{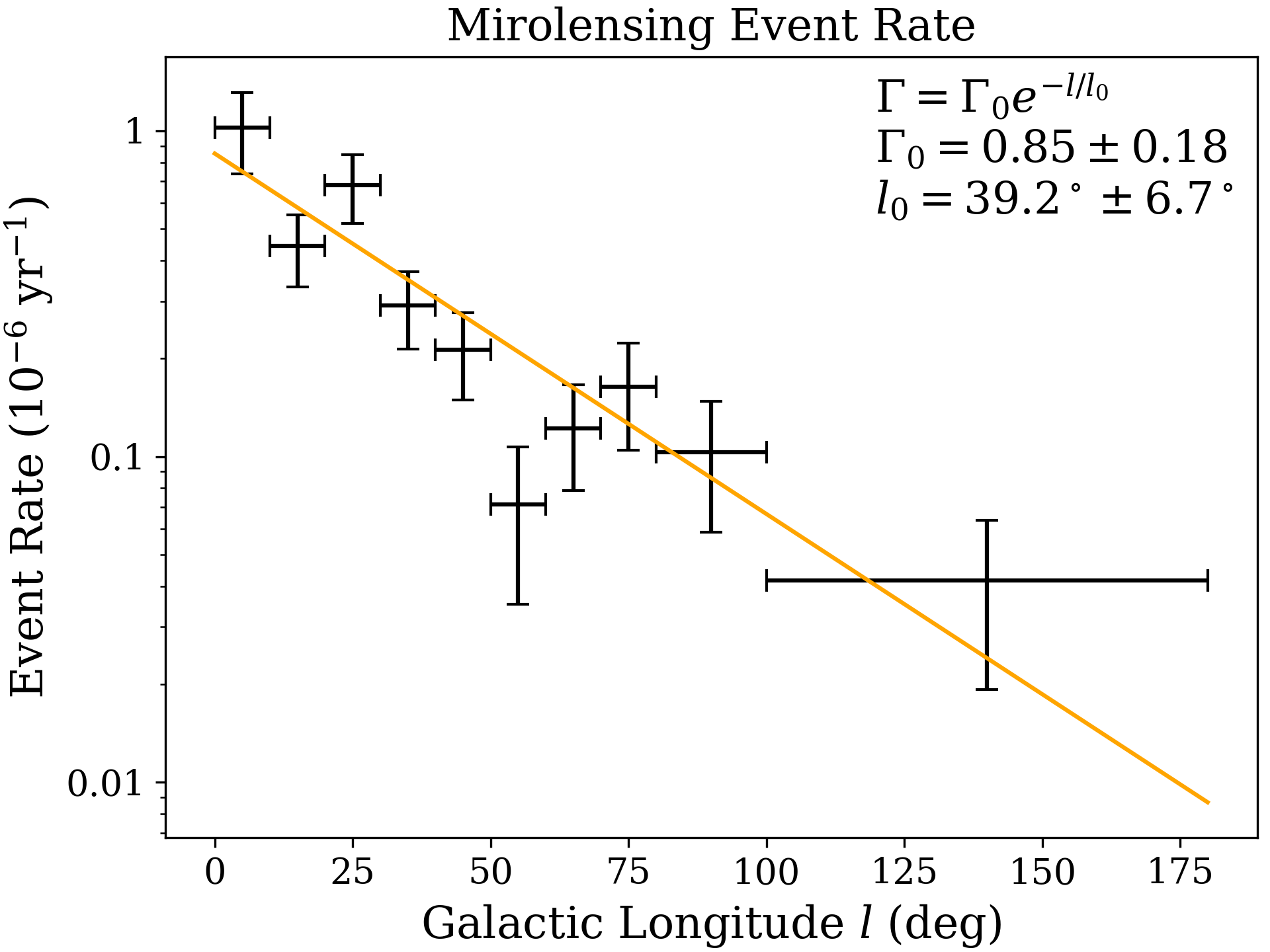}
    \includegraphics[width = 1.0\columnwidth]{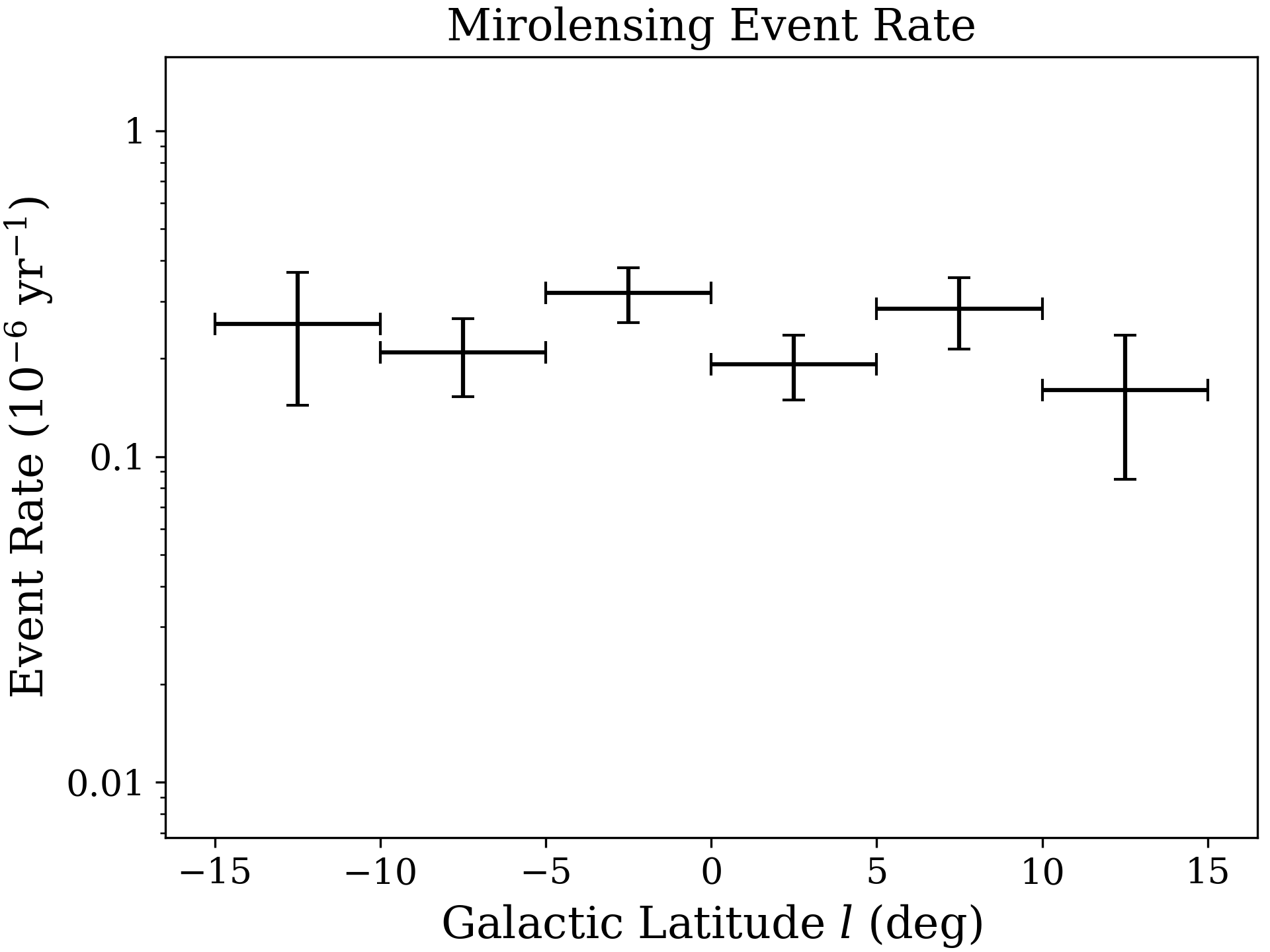}
    \caption{Microlensing optical depth and event rate vs. Galactic longitude and latitude from ZTF DR17 microlensing results. Both optical depth and the event rate drop exponentially with the Galactic longitude while nearly with the Galactic latitude. Their minor fluctuation on the Galactic latitude is similar to results in Figure 6. of \protect\cite{Mroz2020} since their relative value relationship is unaffected by their absolute values.}
    \label{fig:optical-event}
\end{figure*}

The microlensing optical depth is the proportion of sky covered by the angular Einstein rings of microlensing lenses, which is  calculated by \citep{Udalski1994}:
\begin{equation}
    \tau = \frac{\pi}{2 N_\mathrm{src} \Delta T} \sum_{i}\frac{t_{\mathrm{E}, i}}{\epsilon(t_{\mathrm{E}, i})},
\end{equation}
where $N_\mathrm{src}$ is the number of sources in the sky region, and $\Delta T$ is the survey duration: $\Delta T = 1,786\ \mathrm{days}$.

The microlensing event rate measures the rate of the appearance for a source monitored by ZTF:
\begin{equation}
    \Gamma = \frac{1}{N_\mathrm{src} \Delta T}\sum_{i}\frac{1}{\epsilon(t_{\mathrm{E}, i})}.
\end{equation}
The reciprocals of detection efficiencies that appear in the two equations correct the number of microlensing events that our algorithm fails to detect under the current instruments and observation strategies of ZTF.

We calculate the optical depths and event rates at different Galactic coordinates in 7 bins in the Galactic longitude and 6 in the Galactic latitude. Since few events are located in $-180^\circ < l < 0^\circ$ or $|b| > 15^\circ$, we do not calculate optical depths and event rates in those regions. The results are shown in Figure \ref{fig:optical-event}, in which the horizontal error bars manifest the width of bins, while the vertical ones are the errors of measurement defined by Equation (10) in \cite{Udalski1994}. We fit the optical depths and events rates with exponential decreasing models: $\tau,\ \Gamma \propto e^{-l/l_0}$, where $l_0$ is a characteristic length scale. We use the average of Galactic longitudes of fields in each bin to fit. Figure \ref{fig:optical-event} shows the best fit as orange lines. For the optical depth, we find $\tau_0 = (0.19 \pm 0.04)\times 10^{-6},\ l_0 = 39.1^\circ \pm 6.5^\circ$;, while for the event rate, $\Gamma_0 = (0.85 \pm 0.18)\times 10^{-6},\ l_0 = 39.3^\circ \pm 6.7^\circ$. The characteristic length scale $l_0$ agrees well with that calculated by \cite{Mroz2020} in the Southern Galactic plane ($32^\circ$), \cite{Rodriguez2022} in the Northern Galactic plane ($37^\circ$), and the simulations of \cite{Sajadian2019} ($36^\circ$).

Though the trends of the optical depth and event rate with respect to the Galactic coordinates are consistent with results of OGLE \citep{Mroz2020}, the absolute values are discrepant: the OGLE results find $\tau_0 = (0.77 \pm 0.11)\times 10^{-6},\ \Gamma_0 = (3.02 \pm 0.47)\times 10^{-6}$. This discrepancy is likely due to an underestimation of the number of source stars in the ZTF fields. This is due to the stronger blending of sources caused by the larger plate scale of ZTF compared to OGLE and worse seeing at Palomar Observatory compared to Las Campanas Observatory. We conclude that the \textit{relative} trends in event rate and optical depth as a function of Galactic coordinates are consistent between ZTF and OGLE samples, but further work is needed to better account for sources in ZTF fields and then confidently calculate absolute values.

\subsection{Spectroscopy of Possible Events}

For 3 possible events, we obtained optical spectra with the Hale 200-inch telescope at Palomar Observatory, with details given in Appendix \ref{app:1}. Their spectra show no evidence of stellar activity or emission lines, which indicate they are associated with true microlensing events. This also demonstrates the potential of spectroscopy to confirm ZTF microlensing events and remove false positives. Most importantly, this suggests that our high-confidence sample is likely pure and that events we labeled "possible" may be real. 

\subsection{Cross-match with \textit{Gaia} Alerts}

The Cambridge Institute of Astronomy uses \textit{Gaia} data to look for transient sources. The detected signals are classified as \textit{Gaia} Photometric Science Alerts (\textit{Gaia} Alerts; \citealp{Hodgkin2013, Hodgkin2021}). We cross-match all high-confidence and possible events with \textit{Gaia} Alerts \footnote{\url{http://gsaweb.ast.cam.ac.uk/alerts/alertsindex}} within $3^{\prime\prime}$. 124 high-confidence events have 39 matches, and 54 possible events have 7 matches (see Table \ref{parms_high} and \ref{parms_possible}, available online\footnote{\url{https://zenodo.org/doi/10.5281/zenodo.10976973}\label{online_resources}}). \textit{Gaia} Alerts have consistent signals with our observations, and some have already been identified as microlensing candidates, though \textit{Gaia} does not have full coverage on some signals. Since \textit{Gaia} is at Sun-Earth $\text{L}_2$ (0.01 AU from Earth), the parallax signature is not expected to be large for our events since the typical projected Einstein radius on the observer plane is a few AU. 

\subsection{Bayesian Analysis}\label{sec:bayes}

Without observation of the Einstein radius and solid constraint on microlens parallax, we cannot directly measure the physical parameters of lensing systems, such as distances and lens mass. Thus, we conduct Bayesian analysis based on a Galactic model \citep{Yang2021_GalacticModel}. We simulate $10^6$ events for each MCMC solution. We weigh the simulated event $i$ by:
\begin{equation}
    w_{\mathrm{Gal},i} = \Gamma_i \times \mathcal{L}_i(t_\mathrm{E}) \mathcal{L}_i(\bm{\pi}_\mathrm{E}),
\end{equation}
where $ \Gamma_i = \theta_{\mathrm{E},i}\times\mu_{\mathrm{rel},i}$ is the microlensing event rate, and $\mathcal{L}_i(t_\mathrm{E})$ and $\mathcal{L}_i(\bm{\pi}_\mathrm{E})$ are likelihood distributions of $t_\mathrm{E}$ and $\bm{\pi}_\mathrm{E}$ from light curves \citep{Zang2021}.

We summarize the source distance $D_S$, the lens distance $D_L$, the lens mass $M_L$, and the lens-source relative proper motion $\mu_\mathrm{rel}$ in Table \ref{parms_bayesian} for the high-confidence events, available online\footref{online_resources}. The results suggest that all lenses are less massive than the Sun in the Galactic disc. Since we have not measured the Einstein radii, the uncertainties of these values are pretty significant, and the Bayesian analysis results are only for reference.

We search for corresponding sources in the \textit{Gaia} Archive\footnote{\url{https://archives.esac.esa.int/gaia}}, but for most of our sources, the source distances are uncertain since they are too faint. As a result, we cannot improve the accuracy of the Bayesian analysis with \textit{Gaia} data.

\section{Conclusion}\label{sec:5}

We have undertaken a systematic search for microlensing events in the Galactic Plane using 5 years of photometric data from ZTF DR17. We found 124 high-confidence events and 54 possible events, doubling the previous number of microlensing events using only 3 years of ZTF data \citep{Rodriguez2022, 2023medford}. 

We performed an injection-recovery test in ZTF fields, finding events and calculating detection efficiencies. With detection efficiencies as a function of the Einstein timescale, we conclude that the mean $t_E$ of our sample is 51.7$\pm$3.3 days, shorter than that of previous studies of the Galactic plane, but still within $1.5\sigma$

We then binned our events by Galactic latitude and longitude, finding a characteristic length scale of $\sim 39^\circ$ consistent with theoretical predictions and previous surveys. We calculated the microlensing optical depth and event rate as a function of Galactic coordinates. We found that while the trends are similar to that found in OGLE data \citep{Mroz2020}, the overall values are a factor of $\sim 3-4$ lower. We conclude this is likely due to the difficulty of confidently counting all source stars due to blending.

We have shown that ZTF can be used to find microlensing events in the Galactic plane, including those that show a parallax effect and provide an extra constraint on the lens mass. Systematic searches of future ZTF data releases, \textit{Gaia} photometric data, and the upcoming Rubin Observatory Legacy Survey of Space and Time (LSST) will provide a wealth of microlensing events that will shed new light on the population of luminous and dark objects in our Galaxy.

\section{Acknowledgements}
\hfill 

We thank Shrinivas R. Kulkarni for making this work possible. We thank Weicheng Zang and Przemek Mr{\'o}z for their suggestions on developing this paper and our methodology. R.Z. acknowledges support from the National Natural Science Foundation of China (Grant No. 12133005). A. C. R. acknowledges support from an NSF Graduate Research Fellowship. C. Y. L. acknowledges support from a Carnegie Fellowship and a Harrison Fellowship.

We acknowledge the Tsinghua Astrophysics High-Performance Computing platform at Tsinghua University for providing computational and data storage resources that have contributed to the research results reported within this paper. We are grateful to the staff of Palomar Observatory for their work in helping us carry out our observations.  

Based on observations obtained with the Samuel Oschin Telescope 48-inch and the 60-inch Telescope at the Palomar Observatory as part of the Zwicky Transient Facility project. ZTF is supported by the National Science Foundation under Grants No. AST-1440341 and AST-2034437 and a collaboration including current partners Caltech, IPAC, the Weizmann Institute of Science, the Oskar Klein Center at Stockholm University, the University of Maryland, Deutsches Elektronen-Synchrotron and Humboldt University, the TANGO Consortium of Taiwan, the University of Wisconsin at Milwaukee, Trinity College Dublin, Lawrence Livermore National Laboratories, IN2P3, University of Warwick, Ruhr University Bochum, Northwestern University and former partners the University of Washington, Los Alamos National Laboratories, and Lawrence Berkeley National Laboratories. Operations are conducted by COO, IPAC, and UW.

\appendix
\section{Spectroscopic Observations}\label{app:1}

While modeling the photometric signature of microlensing events is often sufficient to confirm their nature, spectroscopy is ultimately the best tool to rule out any false positives. Young star variability and outbursts, dwarf nova outbursts, and Be star outbursts are just some phenomena that may masquerade as microlensing events. Mostly, all these systems and other possible false positives would show emission lines in an optical spectrum. In contrast, the source star in a true microlensing event would be normal. Optical spectroscopy is also helpful since the source properties (e.g., stellar type, luminosity, distance) can be potentially constrained. Precise knowledge of these parameters would allow us to provide additional constraints on the source distance that would help measure the lens mass.

We obtained optical spectroscopy using the Double Spectrograph \citep[DBSP;][]{dbsp} on the Hale Telescope on 14 October 2023. We used the 600/4000 grism on the blue side and the 316/7500 grating on the red side. A 1.0$\arcsec$ slit was used, and the seeing throughout the observation varied between 1.0 -- 1.3$\arcsec$, leading to some slit losses. All P200/DBSP data were reduced with \texttt{DBSP-DRP}\footnote{\url{https://dbsp-drp.readthedocs.io/en/stable/index.html}}, a Python-based pipeline optimized for DBSP built on the more general \texttt{PypeIt} pipeline \citep{2020pypeit}. All fields were sky-subtracted using standard techniques. Internal arc lamps and a standard star for overall flux calibration were used for the wavelength calibration. 

\begin{figure*}
    \centering
    \includegraphics[width = 0.9\linewidth]{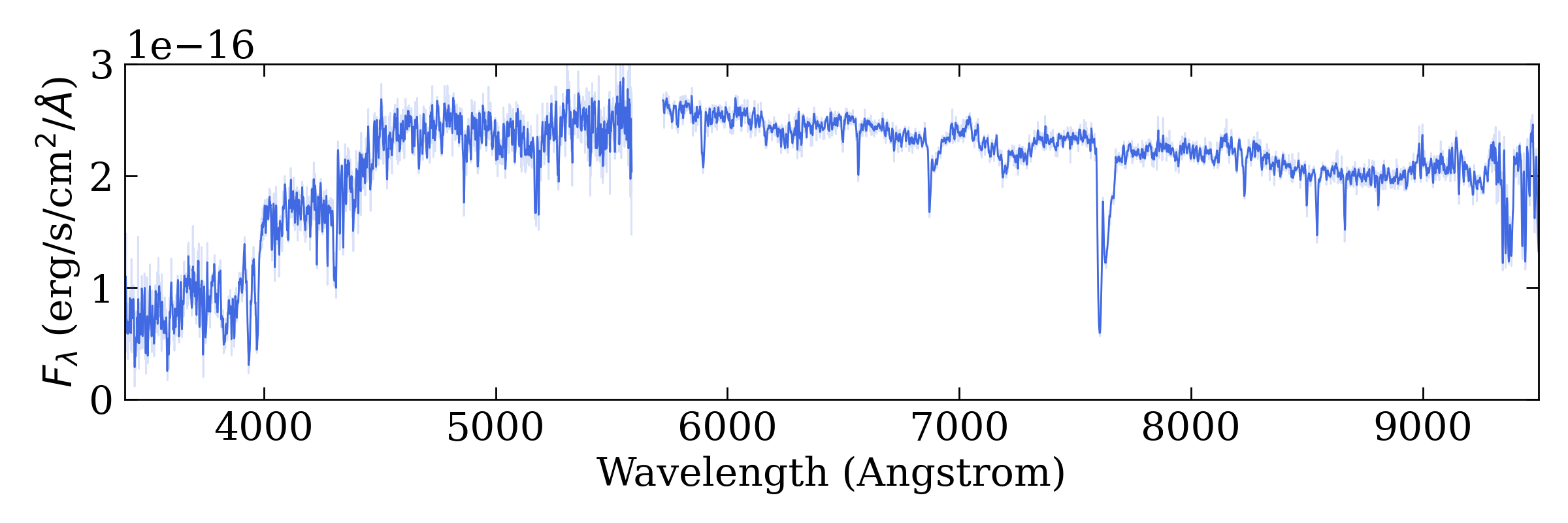}
    \caption{Spectrum of ZTF20acpjazv. No emission lines or indication of stellar activity are present, confirming the source is associated with a high-confidence microlensing event.}
    \label{fig:spectrum}
\end{figure*}

We acquired spectra of three of our microlensing candidates (ZTF18abqbeqv, ZTF20acpjazv, and ZTF22aaajzoi) to check for emission lines. We purposefully chose events whose light curves showed smaller brightening amplitudes than average in our sample but had an excellent fit to the PSPL model. Spectra of all three systems reveal them to be ordinary K and M dwarfs, with no significant ($>3\sigma$) emission lines above the continuum.


\section{Lists of All Events }\label{app:2}

Here we list all 124 high-confidence events and 54 possible events in Table \ref{parms_high} and Table \ref{parms_possible}. In the list of the high-confidence events, we also include parallax model results for events that have. We present the ZTF IDs, cross-matched \textit{Gaia} alerts (if available), coordinates, MCMC model parameters, and $\chi^2/\mathrm{d.o.f}$ in the tables. In the tables, we show HJD'=HJD-2450000 days. Table \ref{parms_high} includes the detection efficiencies of the high-confidence events that are used to calculate the mean Einstein timescale. We provide full lists of events in machine-readable tables, available online \footref{online_resources}. We also show the list of physical parameters of high-confidence events estimated for Bayesian analyses in Table 
\ref{parms_bayesian} (see the Section \ref{sec:bayes}).

We show light curves of 4 events as examples in Figure \ref{fig:examples}. We show one event with a characteristic time scale similar to the mean $t_\mathrm{E}$ of our sample (ZTF19aatwaux) and one event with an exceptionally long timescale (ZTF22aabpdbm). These events have good coverage throughout the baseline as well as the peak. At the same time, we include 2 events identified as false positives that we manually remove from our samples in Figure \ref{fig:false_pos}.

\begin{figure*}[htb]
    \centering
    \includegraphics[width = 0.49\textwidth]{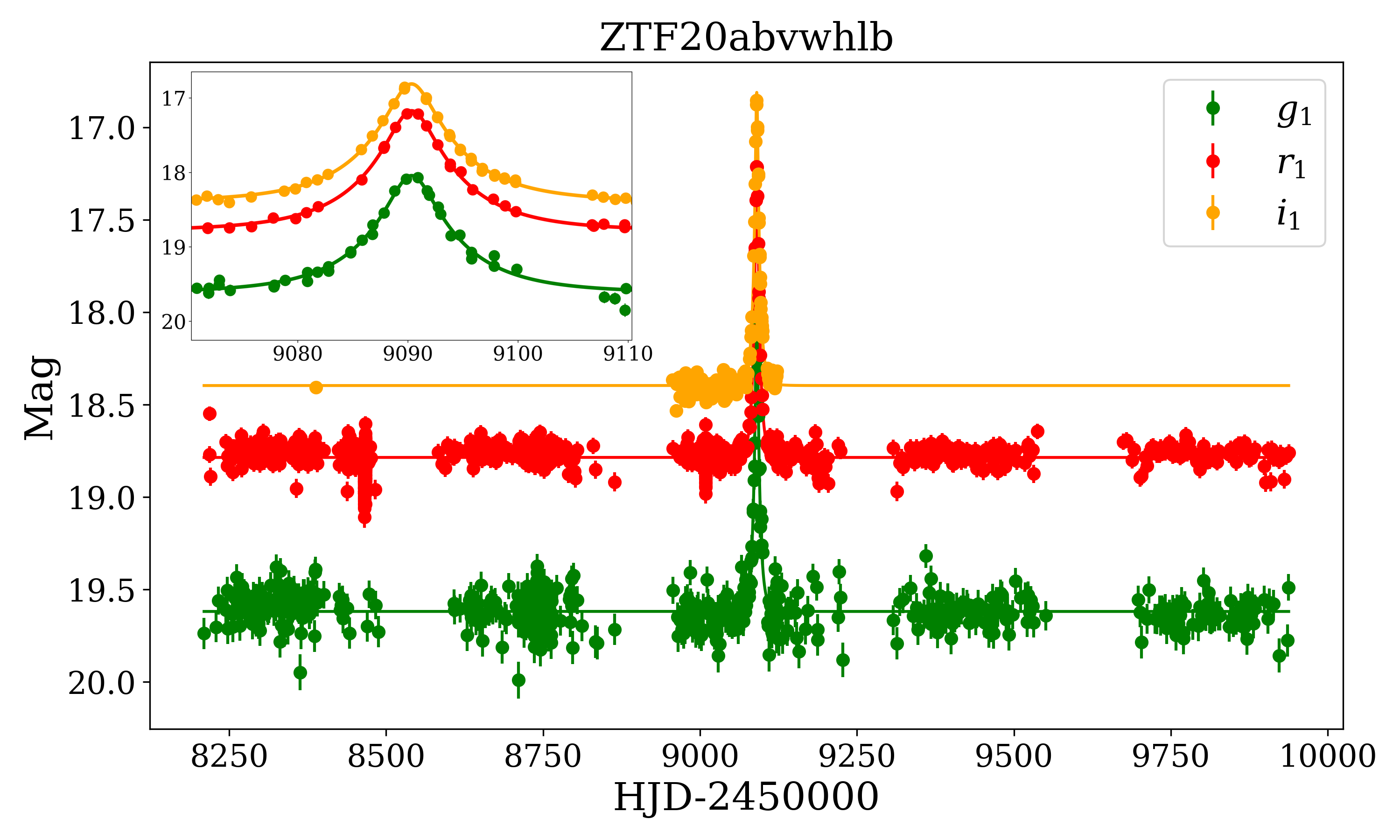}
    \includegraphics[width = 0.49\textwidth]{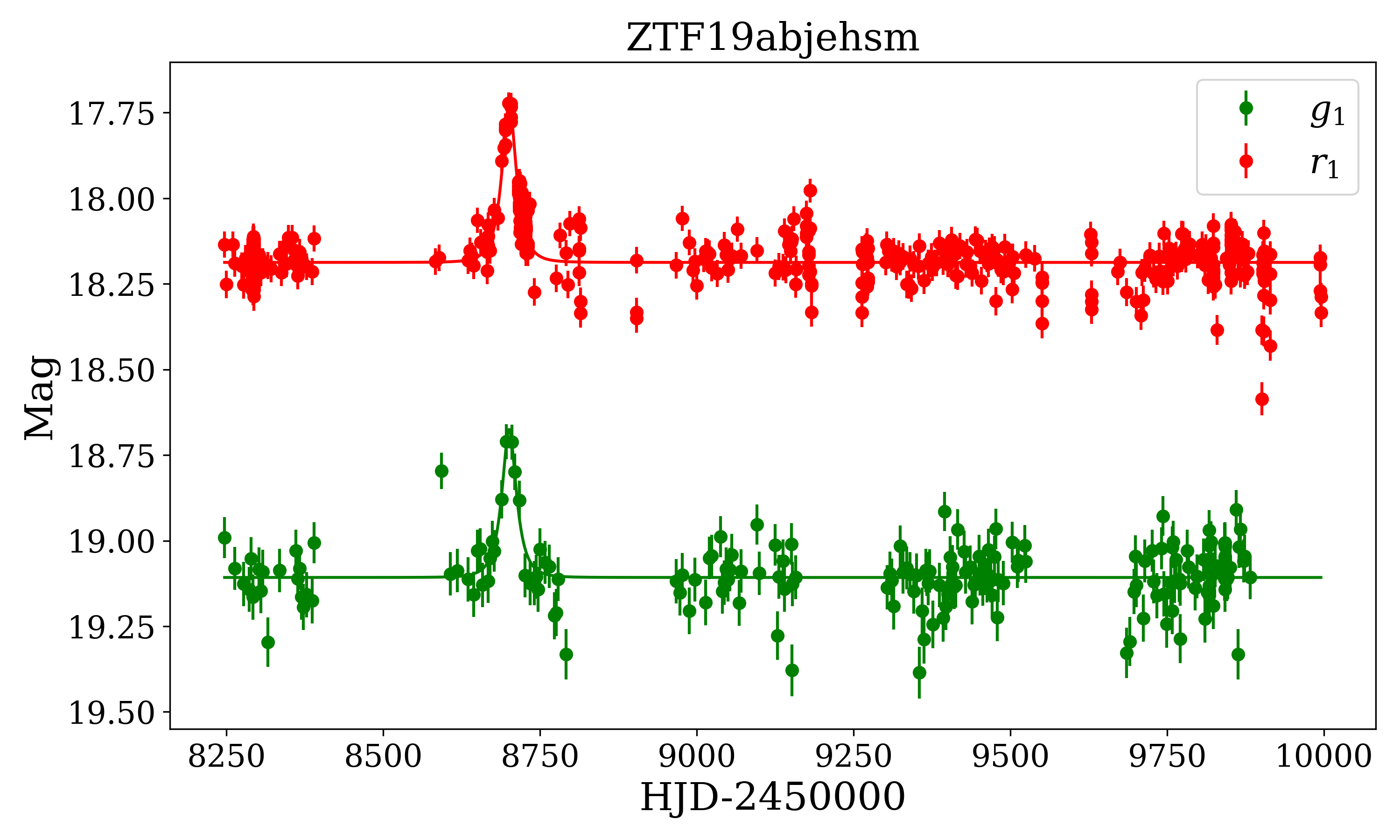}
    \includegraphics[width = 0.49\textwidth]{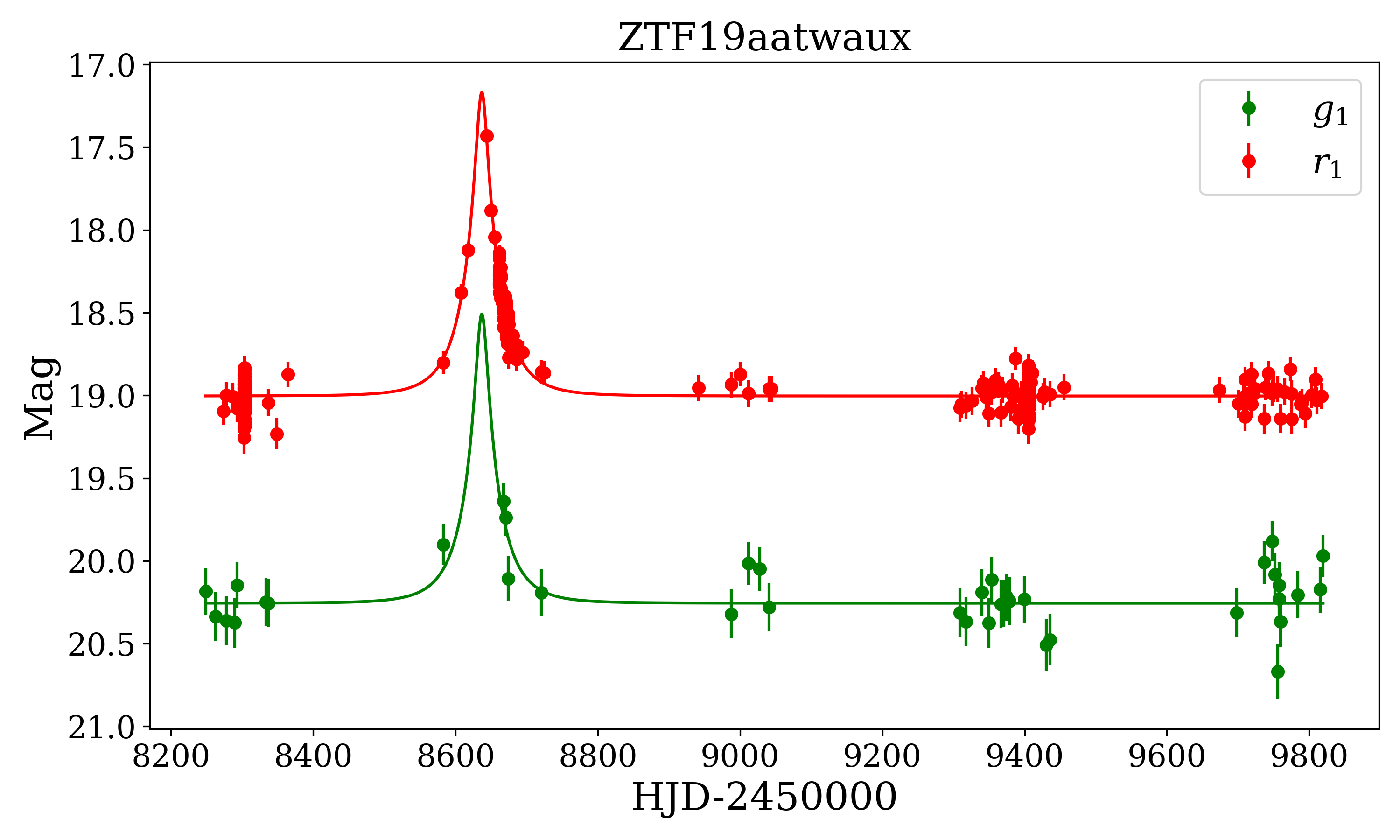}
    \includegraphics[width = 0.49\textwidth]{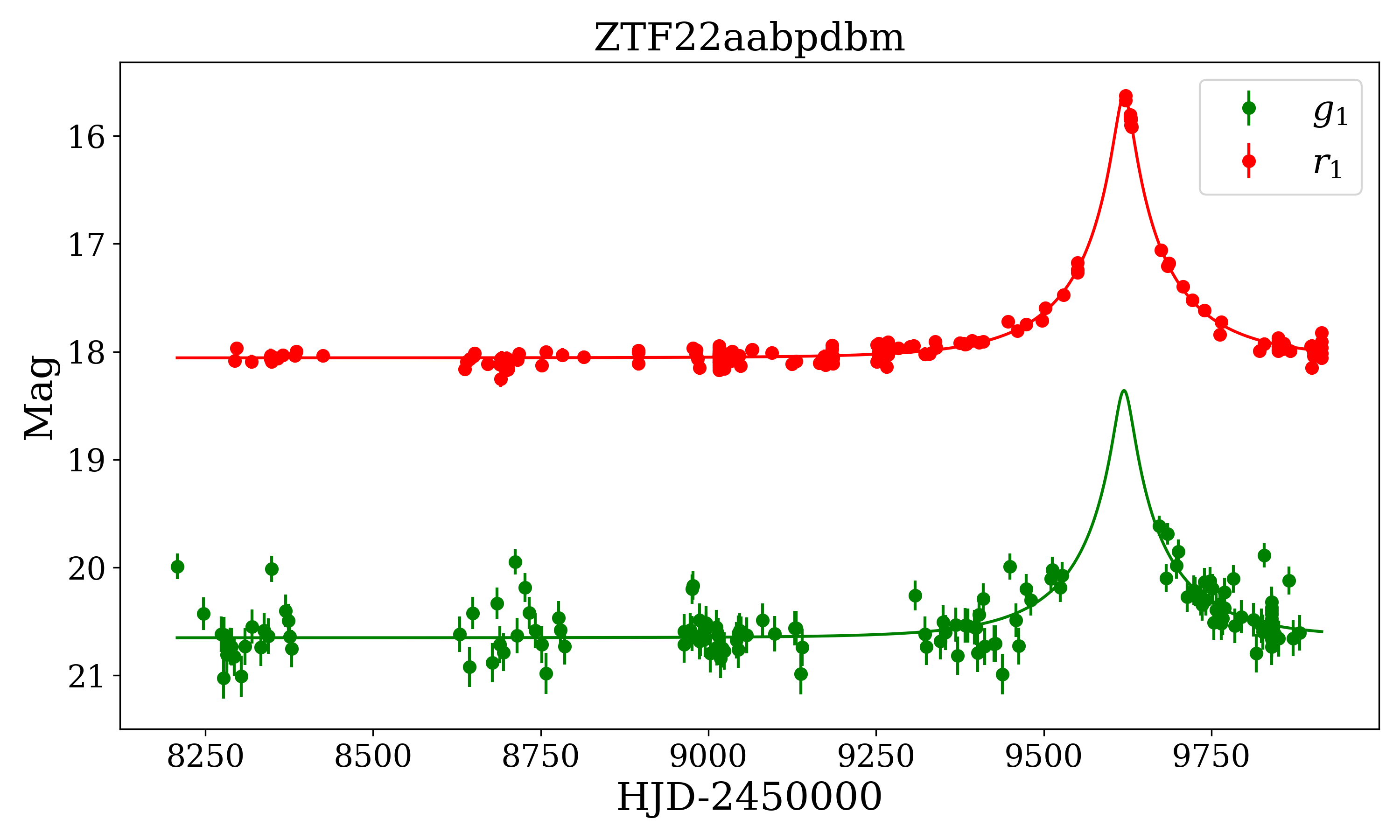}
    \caption{Light curves of ZTF20abvwhlb, ZTF19abjehsm, ZTF19aatwaux, and ZTF22aabpdbm. The Einstein timescale of ZTF20abvwhlb is $8.8^{+0.4}_{-0.3}$ days. The Einstein timescales for the other events are $20.6^{+4.9}_{-3.2}$ days, $51.6^{+7.7}_{-5.2}$ days, and $162.7^{+14.7}_{-13.0}$ days, respectively.}
    \label{fig:examples}
\end{figure*}

\begin{figure*}[htb]
    \centering
    \includegraphics[width = 0.49\textwidth]{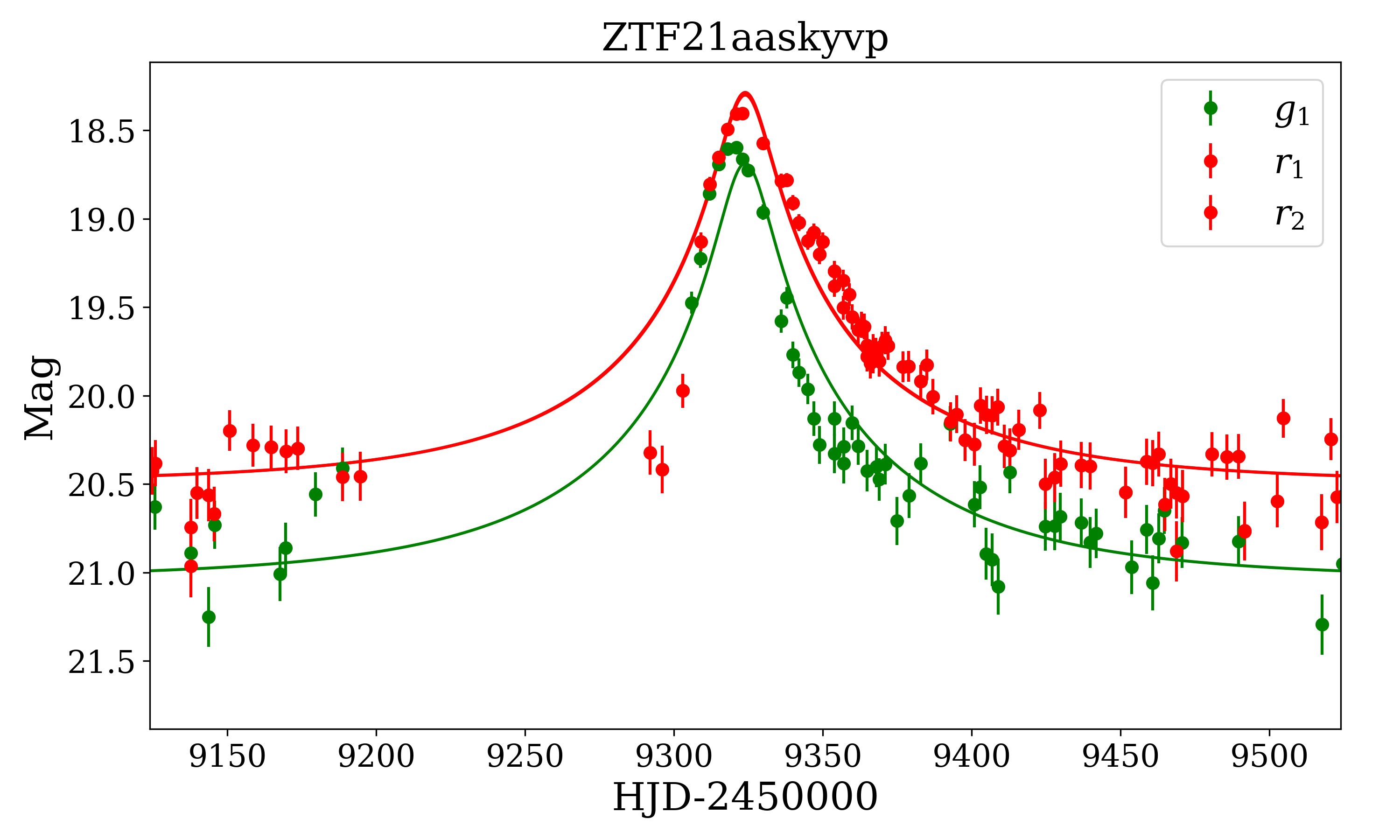}
    \includegraphics[width = 0.49\textwidth]{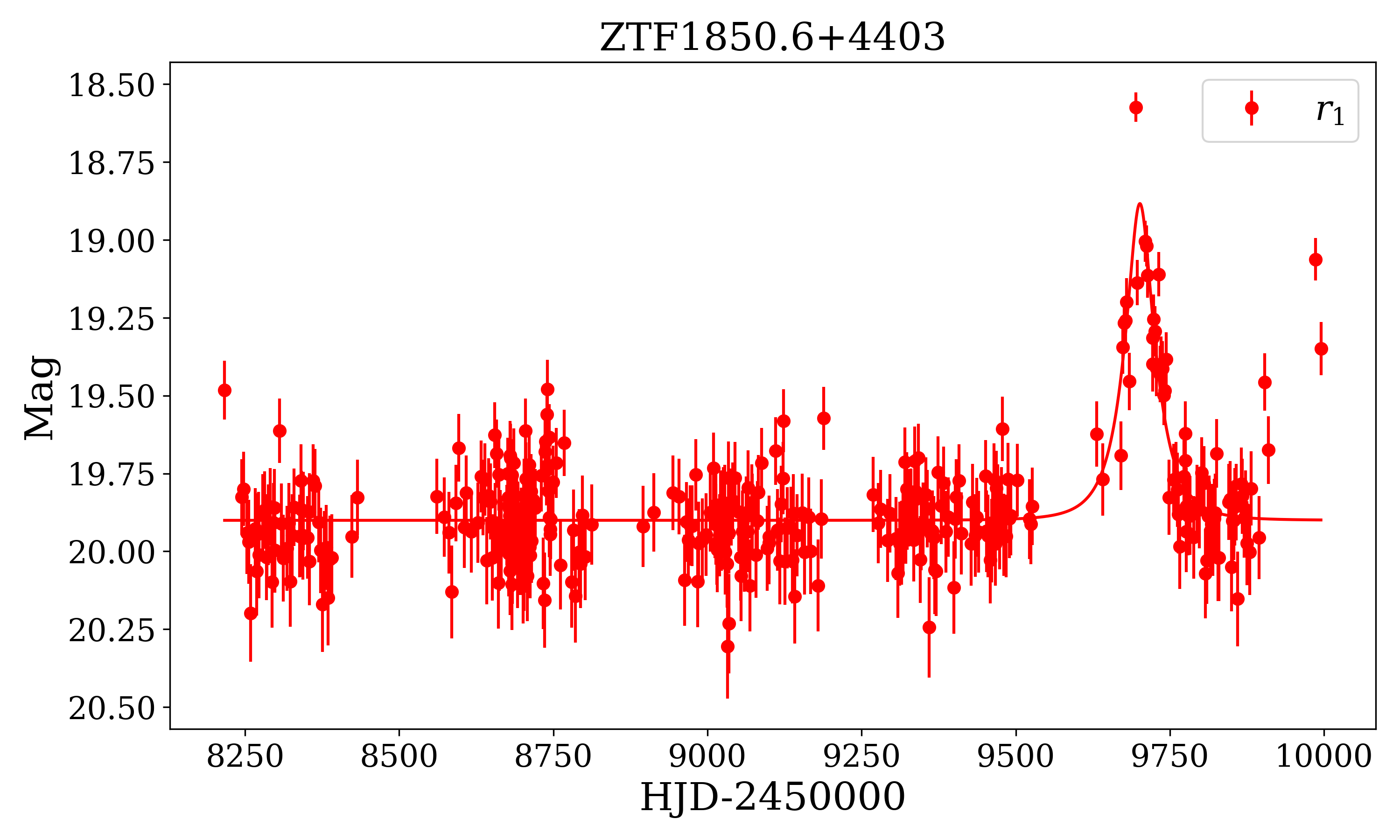}
    \caption{Light curves for two false positive events: ZTF21aaskyvp and ZTF1850.6+4403. ZTF21aaskyvp was classified as a supernova (SN 2021icu, \protect\citealp{ZTF21aaskyvp_report}). We remove it thanks to the exponential decay and color change in the light curve. ZTF1850.6+4403 has a few outliers at the end of its light curve and has a similar shape to many events located far away from it. Hence, we identify it as a false positive.}
    \label{fig:false_pos}
\end{figure*}

\begin{longrotatetable}
    \begin{deluxetable*}{ccccccccccccc}
        \tablecaption{List of high-confidence microlensing events in ZTF DR17 with MCMC Parameters. The upper arrows represent parallax models of their previous events.\label{parms_high}}
        \tablehead{\colhead{ID} & \colhead{\textit{Gaia} Alerts} & \colhead{R.A (J2000)} & \colhead{Decl. (J2000)} & \colhead{$t_0$ (HJD')} & \colhead{$u_0$} & \colhead{$t_\mathrm{E}$} & \colhead{$\pi_\mathrm{E,N}$} & \colhead{$\pi_\mathrm{E,E}$} & \colhead{$r_s$ (mag)} & \colhead{$g_s$ (mag)} & \colhead{$\chi^2$/d.o.f} & \colhead{$\text{det}_\text{eff}$}}
        \startdata
        ZTF20abwxmep & ... & 12.904570 & 60.617889 & $9092.74^{+0.88}_{-0.90}$ & $0.48^{+0.06}_{-0.10}$ & $41.7^{+5.9}_{-3.1}$ & ... & ... & $20.1^{+0.3}_{-0.2}$ & $21.6^{+0.3}_{-0.2}$ & 1229.7/933 & 0.26 \\
        ZTF19adbsiat & ... & 65.329219 & 30.695008 & $8915.88^{+0.72}_{-0.74}$ & $0.68^{+0.04}_{-0.06}$ & $50.4^{+3.1}_{-1.9}$ & ... & ... & $18.3^{+0.2}_{-0.1}$ & $19.3^{+0.2}_{-0.1}$ & 1733.5/988 & 0.29 \\
        ZTF0527.8+5142 & ... & 81.950964 & 51.705438 & $9445.32^{+0.09}_{-0.09}$ & $0.05^{+0.00}_{-0.00}$ & $69.6^{+2.8}_{-2.7}$ & ... & ... & $19.3^{+0.1}_{-0.1}$ & $20.3^{+0.1}_{-0.1}$ & 1536.1/877 & 0.39 \\
        $\uparrow$ & $\uparrow$ & $\uparrow$  & $\uparrow$ & $9443.33^{+0.17}_{-0.11}$ & $-0.00^{+0.00}_{-0.00}$ & $421.0^{+89.5}_{-71.1}$ & $-0.520^{+0.073}_{-0.075}$ & $0.060^{+0.016}_{-0.013}$ & $19.3^{+0.1}_{-0.1}$ & $20.3^{+0.1}_{-0.1}$ & 1416.6/875 & ... \\
        ZTF20abztmco & Gaia20erh & 82.452721 & 15.428445 & $9238.35^{+0.26}_{-0.26}$ & $0.23^{+0.02}_{-0.01}$ & $172.8^{+8.6}_{-8.3}$ & ... & ... & $19.3^{+0.1}_{-0.1}$ & $20.6^{+0.1}_{-0.1}$ & 2582.4/1180 & 0.48 \\
        $\uparrow$ & $\uparrow$ & $\uparrow$  & $\uparrow$ & $9236.34^{+0.58}_{-0.48}$ & $0.15^{+0.25}_{-0.04}$ & $237.2^{+62.4}_{-118.8}$ & $-0.105^{+0.195}_{-0.013}$ & $-0.005^{+0.006}_{-0.102}$ & $19.3^{+0.1}_{-0.1}$ & $20.6^{+0.1}_{-0.1}$ & 2523.1/1178 & ... \\
        ZTF21aadrxrx & ... & 255.942718 & -10.214135 & $9237.50^{+0.09}_{-0.08}$ & $0.06^{+0.00}_{-0.00}$ & $39.2^{+0.9}_{-0.9}$ & ... & ... & $15.7^{+0.0}_{-0.0}$ & $16.9^{+0.0}_{-0.0}$ & 1399.3/1257 & 0.15 \\
        ZTF19aatwaux & ... & 258.208394 & -27.182025 & $8637.02^{+0.56}_{-0.53}$ & $0.15^{+0.04}_{-0.04}$ & $51.6^{+7.7}_{-5.2}$ & ... & ... & $19.3^{+0.3}_{-0.2}$ & $20.6^{+0.3}_{-0.2}$ & 533.1/559 & 0.11 \\
        ZTF21aaqsyeh & ... & 258.385319 & -19.727009 & $9310.94^{+0.35}_{-0.35}$ & $0.40^{+0.02}_{-0.04}$ & $27.6^{+2.0}_{-1.2}$ & ... & ... & $18.5^{+0.2}_{-0.1}$ & $19.4^{+0.2}_{-0.1}$ & 2076.8/711 & 0.09 \\
        ZTF21abdihbw & ... & 258.715905 & -21.966701 & $9382.33^{+0.06}_{-0.06}$ & $0.04^{+0.01}_{-0.00}$ & $45.9^{+4.7}_{-3.9}$ & ... & ... & $20.2^{+0.1}_{-0.1}$ & $21.2^{+0.1}_{-0.1}$ & 857.9/732 & 0.09 \\
        ZTF1721.6-2443 & ... & 260.395064 & -24.721539 & $9770.27^{+0.03}_{-0.02}$ & $0.00^{+0.00}_{-0.00}$ & $41.5^{+3.4}_{-3.0}$ & ... & ... & $19.0^{+0.1}_{-0.1}$ & $20.1^{+0.1}_{-0.1}$ & 746.8/673 & 0.10 \\
        ZTF19abpxurg & ... & 260.994091 & -20.194962 & $8713.86^{+0.05}_{-0.05}$ & $0.05^{+0.01}_{-0.01}$ & $27.7^{+1.8}_{-1.5}$ & ... & ... & $19.4^{+0.1}_{-0.1}$ & $20.6^{+0.1}_{-0.1}$ & 1052.2/880 & 0.09 \\
        ZTF22aaknmsp & ... & 261.417335 & -21.232529 & $9736.85^{+0.21}_{-0.21}$ & $0.11^{+0.01}_{-0.02}$ & $33.1^{+4.3}_{-2.7}$ & ... & ... & $19.7^{+0.2}_{-0.1}$ & $21.5^{+0.2}_{-0.1}$ & 433.4/411 & 0.07 \\
        ZTF22aakolyk & ... & 261.725881 & -11.004709 & $9737.67^{+0.13}_{-0.12}$ & $0.22^{+0.04}_{-0.03}$ & $26.7^{+2.4}_{-2.2}$ & ... & ... & $19.0^{+0.2}_{-0.2}$ & $19.8^{+0.2}_{-0.2}$ & 1106.9/722 & 0.12 \\
        ZTF21aawqeqn & Gaia21auw & 262.325899 & -21.875246 & $9343.11^{+0.53}_{-0.57}$ & $0.14^{+0.03}_{-0.02}$ & $125.2^{+15.2}_{-13.0}$ & ... & ... & $17.8^{+0.2}_{-0.2}$ & $19.6^{+0.2}_{-0.2}$ & 611.6/355 & 0.12 \\
        ZTF21abdikzn & ... & 262.411213 & -25.010295 & $9407.11^{+0.84}_{-0.85}$ & $0.31^{+0.08}_{-0.09}$ & $83.6^{+23.1}_{-12.9}$ & ... & ... & $19.6^{+0.4}_{-0.3}$ & $21.4^{+0.4}_{-0.3}$ & 362.8/298 & 0.12 \\
        ZTF19aazhmab & ... & 264.957266 & -20.894358 & $8642.29^{+0.11}_{-0.14}$ & $0.03^{+0.02}_{-0.02}$ & $82.7^{+14.8}_{-11.8}$ & ... & ... & $19.7^{+0.3}_{-0.3}$ & $20.7^{+0.2}_{-0.2}$ & 94.3/85 & 0.11 \\
        ZTF21abehdtd & ... & 265.277131 & -22.024187 & $9375.80^{+0.45}_{-0.47}$ & $0.24^{+0.02}_{-0.03}$ & $79.4^{+7.2}_{-4.1}$ & ... & ... & $18.6^{+0.2}_{-0.1}$ & $20.3^{+0.2}_{-0.1}$ & 344.3/280 & 0.11 \\
        ZTF22aazwyen & ... & 266.027068 & -19.565755 & $9801.27^{+0.58}_{-0.53}$ & $0.49^{+0.15}_{-0.16}$ & $21.6^{+5.7}_{-3.2}$ & ... & ... & $18.1^{+0.6}_{-0.4}$ & $19.7^{+0.6}_{-0.4}$ & 1108.7/1250 & 0.07 \\
        ZTF1752.3-1551 & Gaia22brc & 268.066416 & -15.860507 & $9682.24^{+0.16}_{-0.13}$ & $0.04^{+0.03}_{-0.03}$ & $40.4^{+13.8}_{-10.0}$ & ... & ... & $19.7^{+0.4}_{-0.4}$ & $20.7^{+0.4}_{-0.4}$ & 1088.6/503 & 0.15 \\
        ZTF21aavhwyj & ... & 268.178174 & -15.623007 & $9352.87^{+0.11}_{-0.12}$ & $0.14^{+0.01}_{-0.01}$ & $47.4^{+2.0}_{-1.8}$ & ... & ... & $17.6^{+0.1}_{-0.1}$ & $18.7^{+0.1}_{-0.1}$ & 410.3/520 & 0.16 \\
        ZTF1753.1-2006 & ... & 268.285211 & -20.114489 & $8266.61^{+0.56}_{-0.59}$ & $0.16^{+0.05}_{-0.05}$ & $40.0^{+9.2}_{-5.6}$ & ... & ... & $18.5^{+0.4}_{-0.3}$ & $21.0^{+0.4}_{-0.3}$ & 269.9/345 & 0.15 \\
        ZTF20aaukggm & ... & 268.528270 & -20.022328 & $8962.00^{+1.13}_{-1.61}$ & $0.11^{+0.02}_{-0.02}$ & $53.5^{+6.3}_{-5.6}$ & ... & ... & $18.2^{+0.3}_{-0.3}$ & $20.6^{+0.2}_{-0.3}$ & 614.9/401 & 0.19 \\
        ZTF20abqkcsf & ... & 268.552367 & -1.281956 & $9078.65^{+1.80}_{-1.92}$ & $0.14^{+0.05}_{-0.05}$ & $55.5^{+21.5}_{-9.6}$ & ... & ... & $20.9^{+0.6}_{-0.4}$ & $21.9^{+0.6}_{-0.4}$ & 642.2/440 & 0.24 \\
        ZTF21abehnjv & ... & 269.186521 & -17.938606 & $9399.29^{+0.19}_{-0.18}$ & $0.22^{+0.02}_{-0.02}$ & $48.8^{+3.7}_{-3.0}$ & ... & ... & $18.4^{+0.1}_{-0.1}$ & $20.1^{+0.1}_{-0.1}$ & 622.2/504 & 0.16 \\
        ZTF20abkynur & ... & 269.342506 & -16.534367 & $9048.38^{+0.13}_{-0.12}$ & $0.14^{+0.01}_{-0.01}$ & $17.2^{+1.2}_{-1.1}$ & ... & ... & $19.7^{+0.1}_{-0.1}$ & $21.5^{+0.1}_{-0.1}$ & 463.2/236 & 0.08 \\
        ZTF22abcflpn & ... & 270.864163 & -12.808809 & $9821.76^{+0.17}_{-0.16}$ & $0.08^{+0.02}_{-0.02}$ & $28.8^{+4.2}_{-2.9}$ & ... & ... & $20.2^{+0.2}_{-0.2}$ & $22.6^{+0.2}_{-0.2}$ & 768.4/370 & 0.16 \\
        \vdots & \vdots & \vdots & \vdots & \vdots & \vdots & \vdots & \vdots & \vdots & \vdots & \vdots & \vdots \\
        \enddata
    \end{deluxetable*}
\end{longrotatetable}

\begin{longrotatetable}
    \begin{deluxetable*}{cccccccccccc}
        \tablecaption{List of possible microlensing events in ZTF DR17 with MCMC Parameters.\label{parms_possible}}
        \tablehead{\colhead{ID} & \colhead{\textit{Gaia} Alerts} & \colhead{R.A (J2000)} & \colhead{Decl. (J2000)} & \colhead{$t_0$ (HJD')} & \colhead{$u_0$} & \colhead{$t_\mathrm{E}$} & \colhead{$\pi_\mathrm{E,N}$} & \colhead{$\pi_\mathrm{E,E}$} & \colhead{$r_s$ (mag)} & \colhead{$g_s$ (mag)} & \colhead{$\chi^2$/d.o.f}}
        \startdata
        ZTF19abfrdeg & ... & 17.324452 & 54.493693 & $8307.49^{+0.69}_{-0.71}$ & $0.21^{+0.12}_{-0.09}$ & $37.1^{+19.7}_{-10.1}$ & ... & ... & $21.4^{+0.7}_{-0.6}$ & $22.7^{+0.7}_{-0.6}$ & 349.9/189 \\
        ZTF0139.2+5403 & ... & 24.793413 & 54.063825 & $8296.18^{+0.21}_{-0.20}$ & $0.03^{+0.04}_{-0.02}$ & $24.3^{+5.4}_{-4.3}$ & ... & ... & $21.1^{+0.3}_{-0.3}$ & $23.1^{+0.2}_{-0.3}$ & 1441.7/613 \\
        ZTF18acdyfuk & ... & 37.180587 & 38.914464 & $8441.71^{+0.68}_{-0.66}$ & $0.23^{+0.03}_{-0.03}$ & $45.4^{+4.9}_{-4.1}$ & ... & ... & $21.0^{+0.2}_{-0.2}$ & $21.0^{+0.2}_{-0.2}$ & 176.8/106 \\
        ZTF19aabbuqn & ... & 48.694297 & 62.343460 & $8506.15^{+0.44}_{-0.43}$ & $0.22^{+0.10}_{-0.07}$ & $45.2^{+14.0}_{-10.2}$ & ... & ... & $20.1^{+0.5}_{-0.5}$ & $21.6^{+0.4}_{-0.5}$ & 656.3/270 \\
        ZTF18acbvsfm & ... & 55.872634 & 39.674278 & $8418.93^{+1.64}_{-1.86}$ & $0.14^{+0.03}_{-0.03}$ & $91.6^{+18.5}_{-12.0}$ & ... & ... & $21.1^{+0.3}_{-0.2}$ & $21.4^{+0.3}_{-0.2}$ & 103.3/61 \\
        ZTF0354.4+2948 & ... & 58.590351 & 29.802082 & $8450.39^{+2.42}_{-2.53}$ & $0.16^{+0.06}_{-0.06}$ & $206.7^{+85.0}_{-42.1}$ & ... & ... & $20.9^{+0.5}_{-0.4}$ & ... & 207.9/69 \\
        ZTF20actffyp & ... & 62.413830 & 44.068973 & $9241.47^{+1.16}_{-1.18}$ & $0.46^{+0.19}_{-0.12}$ & $115.1^{+25.8}_{-22.7}$ & ... & ... & $19.6^{+0.5}_{-0.6}$ & $20.5^{+0.5}_{-0.6}$ & 1687.0/1097 \\
        ZTF20aacghde & ... & 66.840857 & 17.220539 & $8860.68^{+0.79}_{-0.77}$ & $0.46^{+0.12}_{-0.15}$ & $19.2^{+5.7}_{-2.8}$ & ... & ... & $20.1^{+0.6}_{-0.4}$ & $21.1^{+0.6}_{-0.4}$ & 985.3/403 \\
        ZTF0545.5+5952 & ... & 86.376567 & 59.880607 & $8840.60^{+0.35}_{-0.36}$ & $0.32^{+0.05}_{-0.06}$ & $29.0^{+4.3}_{-2.8}$ & ... & ... & $20.3^{+0.3}_{-0.2}$ & $20.9^{+0.3}_{-0.2}$ & 511.4/269 \\
        ZTF19abpangr & ... & 90.818465 & 53.608017 & $8698.10^{+2.71}_{-2.24}$ & $0.04^{+0.01}_{-0.02}$ & $109.2^{+20.2}_{-13.7}$ & ... & ... & $21.0^{+0.3}_{-0.3}$ & $22.2^{+0.3}_{-0.3}$ & 1048.0/568 \\
        ZTF20acvhjbw & ... & 100.236340 & 47.608269 & $9191.09^{+0.44}_{-0.45}$ & $0.45^{+0.11}_{-0.11}$ & $31.0^{+6.7}_{-4.3}$ & ... & ... & $20.6^{+0.4}_{-0.4}$ & $21.1^{+0.4}_{-0.4}$ & 1687.0/858 \\
        ZTF20abwiild & Gaia20dzx & 108.819464 & 16.029165 & $9089.08^{+0.41}_{-0.43}$ & $0.11^{+0.01}_{-0.01}$ & $127.3^{+7.8}_{-6.8}$ & ... & ... & $20.3^{+0.1}_{-0.1}$ & $20.9^{+0.1}_{-0.1}$ & 1335.0/610 \\
        ZTF170.4-2209 & ... & 255.107688 & -22.149985 & $8667.65^{+0.04}_{-0.04}$ & $0.02^{+0.01}_{-0.01}$ & $12.3^{+4.4}_{-2.7}$ & ... & ... & $22.0^{+0.4}_{-0.3}$ & $21.8^{+0.4}_{-0.3}$ & 786.8/325 \\
        ZTF22aakboak & ... & 266.033334 & -16.731047 & $9718.29^{+0.29}_{-0.16}$ & $0.00^{+0.00}_{-0.00}$ & $22662.7^{+872921.3}_{-21381.0}$ & ... & ... & $28.3^{+4.0}_{-3.1}$ & ... & 131.7/119 \\
        ZTF22abcflaq & ... & 267.861895 & -12.370507 & $9843.67^{+0.93}_{-0.91}$ & $0.58^{+0.04}_{-0.06}$ & $58.1^{+4.1}_{-2.7}$ & ... & ... & $19.3^{+0.2}_{-0.1}$ & $20.9^{+0.2}_{-0.1}$ & 1190.4/781 \\
        ZTF1752.5-2006 & ... & 268.131417 & -20.102409 & $9433.10^{+0.61}_{-0.48}$ & $0.24^{+0.07}_{-0.08}$ & $20.2^{+4.8}_{-2.9}$ & ... & ... & $19.9^{+0.4}_{-0.3}$ & ... & 412.4/282 \\
        ZTF1756.4-1615 & ... & 269.099954 & -16.256962 & $8612.39^{+151.07}_{-4.83}$ & $0.69^{+0.36}_{-0.17}$ & $131.2^{+44.7}_{-14.9}$ & ... & ... & $18.6^{+0.5}_{-0.3}$ & $20.3^{+0.5}_{-0.3}$ & 403.2/481 \\
        ZTF21aaygooe & ... & 269.206782 & -18.039390 & $9355.48^{+1.33}_{-1.45}$ & $0.30^{+0.04}_{-0.06}$ & $48.7^{+7.5}_{-4.5}$ & ... & ... & $18.8^{+0.3}_{-0.2}$ & $20.7^{+0.3}_{-0.2}$ & 183.1/195 \\
        ZTF22abcfjfp & ... & 271.174005 & -18.458122 & $9823.26^{+0.27}_{-0.25}$ & $0.40^{+0.03}_{-0.05}$ & $14.1^{+1.3}_{-0.8}$ & ... & ... & $19.0^{+0.2}_{-0.1}$ & ... & 592.0/451 \\
        ZTF18ablrdcc & Gaia18chq & 271.439134 & -12.014536 & $8353.77^{+0.46}_{-0.50}$ & $0.17^{+0.02}_{-0.03}$ & $50.9^{+5.0}_{-3.9}$ & ... & ... & $20.1^{+0.2}_{-0.1}$ & $22.3^{+0.2}_{-0.1}$ & 445.1/357 \\
        ZTF19aawchkq & ... & 271.794671 & -14.315653 & $8634.69^{+1.27}_{-1.24}$ & $0.11^{+0.06}_{-0.04}$ & $91.8^{+38.3}_{-23.0}$ & ... & ... & $21.5^{+0.5}_{-0.5}$ & $22.9^{+0.5}_{-0.5}$ & 733.5/577 \\
        ZTF18ablrbkj & ... & 271.850478 & -10.314385 & $8260.99^{+1.15}_{-1.19}$ & $0.12^{+0.07}_{-0.05}$ & $111.6^{+54.5}_{-34.4}$ & ... & ... & $21.9^{+0.6}_{-0.6}$ & ... & 335.2/256 \\
        ZTF188.2+2207 & ... & 272.062138 & 22.128554 & $9168.08^{+0.49}_{-0.49}$ & $0.18^{+0.03}_{-0.04}$ & $42.8^{+9.0}_{-5.5}$ & ... & ... & $20.9^{+0.3}_{-0.2}$ & $21.2^{+0.3}_{-0.2}$ & 411.3/239 \\
        ZTF18abaphdu & ... & 275.646637 & -0.018457 & $8259.21^{+0.11}_{-0.12}$ & $0.14^{+0.03}_{-0.03}$ & $15.4^{+3.4}_{-2.3}$ & ... & ... & $20.9^{+0.3}_{-0.2}$ & ... & 186.3/114 \\
        ZTF18absrimp & Gaia18cnm & 275.878861 & -7.332711 & $8394.80^{+8.12}_{-11.22}$ & $0.17^{+0.16}_{-0.12}$ & $65.8^{+23.3}_{-9.8}$ & ... & ... & $19.6^{+0.9}_{-0.5}$ & ... & 646.3/551 \\
        ZTF1829.7-1829 & ... & 277.436171 & -18.499529 & $8681.57^{+1.23}_{-1.27}$ & $0.07^{+0.17}_{-0.05}$ & $322.4^{+912.9}_{-207.5}$ & ... & ... & $21.7^{+1.6}_{-1.5}$ & $23.1^{+1.6}_{-1.5}$ & 298.2/344 \\
        ZTF1836.3+0354 & ... & 279.083032 & 3.904392 & $9783.47^{+0.85}_{-0.84}$ & $0.27^{+0.09}_{-0.11}$ & $66.5^{+31.2}_{-12.8}$ & ... & ... & $21.3^{+0.7}_{-0.4}$ & ... & 212.4/166 \\
        \vdots & \vdots & \vdots & \vdots & \vdots & \vdots & \vdots & \vdots & \vdots & \vdots & \vdots & \vdots \\
        \enddata
    \end{deluxetable*}
\end{longrotatetable}

\startlongtable
\begin{deluxetable*}{cccccc}
    \tablecaption{List of physical parameters of high-confidence events from Bayesian analyses.\label{parms_bayesian}}
    \tablehead{\colhead{ID} & \colhead{Model} & \colhead{$D_S$ (kpc)} & \colhead{$D_L$ (kpc)} & \colhead{$M_L$ ($M_\odot$)} & \colhead{$\mu_\mathrm{rel}$ ($\mathrm{mas\ yr^{-1}}$)}}
    \startdata
    ZTF20abwxmep & no parallax & $2.33_{+0.79}^{-0.91}$ & $0.91_{+0.79}^{-0.51}$ & $0.36_{+0.37}^{-0.22}$ & $10.40_{+12.88}^{-5.53}$ \\
    ZTF19adbsiat & no parallax & $1.62_{+0.93}^{-0.72}$ & $0.55_{+0.53}^{-0.30}$ & $0.47_{+0.40}^{-0.28}$ & $12.83_{+15.33}^{-7.00}$ \\
    ZTF0527.8+5142 & no parallax & $1.98_{+0.90}^{-0.83}$ & $0.70_{+0.61}^{-0.37}$ & $0.55_{+0.39}^{-0.31}$ & $8.39_{+9.89}^{-4.45}$ \\
    $\uparrow$ & parallax ($u_0<0$) & $2.16_{+0.75}^{-0.96}$ & $0.73_{+0.49}^{-0.36}$ & $0.54_{+0.45}^{-0.21}$ & $5.80_{+6.29}^{-2.61}$ \\
    ZTF20abztmco & no parallax & $2.15_{+0.86}^{-0.91}$ & $0.89_{+0.61}^{-0.48}$ & $0.74_{+0.44}^{-0.36}$ & $3.74_{+6.75}^{-2.21}$ \\
    $\uparrow$ & parallax ($u_0>0$) & $2.13_{+0.88}^{-0.87}$ & $1.47_{+0.84}^{-0.59}$ & $0.85_{+0.42}^{-0.32}$ & $3.67_{+5.04}^{-2.01}$ \\
    ZTF21aadrxrx & no parallax & $2.12_{+0.91}^{-0.90}$ & $0.77_{+0.76}^{-0.45}$ & $0.42_{+0.35}^{-0.25}$ & $12.25_{+14.08}^{-6.21}$ \\
    ZTF19aatwaux & no parallax & $2.79_{+0.52}^{-0.87}$ & $1.28_{+0.83}^{-0.70}$ & $0.43_{+0.37}^{-0.26}$ & $7.68_{+8.52}^{-3.64}$ \\
    ZTF21aaqsyeh & no parallax & $2.45_{+0.74}^{-0.98}$ & $1.02_{+0.95}^{-0.64}$ & $0.24_{+0.31}^{-0.14}$ & $11.10_{+16.51}^{-5.50}$ \\
    ZTF21abdihbw & no parallax & $2.65_{+0.61}^{-0.91}$ & $1.15_{+0.85}^{-0.65}$ & $0.40_{+0.36}^{-0.24}$ & $8.60_{+10.03}^{-4.13}$ \\
    ZTF1721.6-2443 & no parallax & $2.76_{+0.53}^{-0.88}$ & $1.29_{+0.86}^{-0.73}$ & $0.37_{+0.35}^{-0.22}$ & $8.28_{+9.90}^{-3.91}$ \\
    ZTF19abpxurg & no parallax & $2.59_{+0.65}^{-0.97}$ & $1.16_{+0.94}^{-0.72}$ & $0.24_{+0.30}^{-0.14}$ & $10.29_{+14.88}^{-5.01}$ \\
    ZTF22aaknmsp & no parallax & $2.68_{+0.60}^{-0.94}$ & $1.23_{+0.92}^{-0.73}$ & $0.29_{+0.33}^{-0.17}$ & $9.30_{+12.14}^{-4.43}$ \\
    ZTF22aakolyk & no parallax & $2.30_{+0.82}^{-0.96}$ & $0.91_{+0.91}^{-0.57}$ & $0.25_{+0.31}^{-0.15}$ & $12.40_{+18.53}^{-6.21}$ \\
    ZTF21aawqeqn & no parallax & $2.88_{+0.46}^{-0.78}$ & $1.27_{+0.71}^{-0.60}$ & $0.66_{+0.42}^{-0.34}$ & $4.14_{+4.22}^{-2.03}$ \\
    ZTF21abdikzn & no parallax & $2.88_{+0.46}^{-0.80}$ & $1.34_{+0.79}^{-0.69}$ & $0.51_{+0.40}^{-0.30}$ & $6.10_{+6.65}^{-2.93}$ \\
    ZTF19aazhmab & no parallax & $2.89_{+0.45}^{-0.78}$ & $1.31_{+0.76}^{-0.66}$ & $0.55_{+0.40}^{-0.31}$ & $5.77_{+6.17}^{-2.74}$ \\
    ZTF21abehdtd & no parallax & $2.93_{+0.42}^{-0.77}$ & $1.35_{+0.76}^{-0.67}$ & $0.56_{+0.39}^{-0.31}$ & $5.53_{+5.53}^{-2.58}$ \\
    ZTF22aazwyen & no parallax & $2.74_{+0.55}^{-0.95}$ & $1.40_{+0.97}^{-0.89}$ & $0.16_{+0.26}^{-0.09}$ & $9.68_{+13.73}^{-4.44}$ \\
    ZTF1752.3-1551 & no parallax & $2.78_{+0.52}^{-0.89}$ & $1.36_{+0.89}^{-0.78}$ & $0.30_{+0.36}^{-0.20}$ & $8.47_{+10.11}^{-3.98}$ \\
    ZTF21aavhwyj & no parallax & $2.82_{+0.49}^{-0.83}$ & $1.33_{+0.83}^{-0.72}$ & $0.43_{+0.36}^{-0.25}$ & $7.62_{+7.93}^{-3.57}$ \\
    ZTF1753.1-2006 & no parallax & $2.89_{+0.44}^{-0.82}$ & $1.49_{+0.86}^{-0.82}$ & $0.33_{+0.35}^{-0.21}$ & $7.67_{+8.54}^{-3.49}$ \\
    ZTF20aaukggm & no parallax & $2.93_{+0.43}^{-0.77}$ & $1.46_{+0.80}^{-0.76}$ & $0.45_{+0.37}^{-0.26}$ & $6.67_{+6.86}^{-3.04}$ \\
    ZTF20abqkcsf & no parallax & $2.43_{+0.74}^{-0.97}$ & $0.98_{+0.84}^{-0.58}$ & $0.36_{+0.40}^{-0.24}$ & $10.14_{+13.77}^{-5.13}$ \\
    ZTF21abehnjv & no parallax & $2.91_{+0.43}^{-0.80}$ & $1.44_{+0.82}^{-0.76}$ & $0.43_{+0.37}^{-0.25}$ & $7.03_{+7.34}^{-3.20}$ \\
    ZTF20abkynur & no parallax & $2.77_{+0.54}^{-0.93}$ & $1.50_{+0.97}^{-0.98}$ & $0.13_{+0.20}^{-0.07}$ & $9.80_{+13.89}^{-4.33}$ \\
    ZTF22abcflpn & no parallax & $2.79_{+0.52}^{-0.88}$ & $1.34_{+0.93}^{-0.78}$ & $0.28_{+0.33}^{-0.17}$ & $9.30_{+10.71}^{-4.34}$ \\
    ZTF19abbwpls & no parallax & $2.91_{+0.43}^{-0.78}$ & $1.47_{+0.87}^{-0.80}$ & $0.36_{+0.36}^{-0.22}$ & $7.85_{+7.55}^{-3.54}$ \\
    ZTF19abagvae & no parallax & $3.03_{+0.35}^{-0.68}$ & $1.50_{+0.70}^{-0.61}$ & $0.70_{+0.42}^{-0.35}$ & $3.05_{+2.51}^{-1.40}$ \\
    ZTF19aaonska & no parallax & $2.78_{+0.52}^{-0.82}$ & $1.26_{+0.76}^{-0.61}$ & $0.55_{+0.40}^{-0.31}$ & $6.39_{+4.72}^{-2.86}$ \\
    ZTF21abihdde & no parallax & $2.99_{+0.38}^{-0.72}$ & $1.50_{+0.76}^{-0.70}$ & $0.55_{+0.44}^{-0.34}$ & $4.96_{+5.07}^{-2.51}$ \\
    ZTF21abrgqfr & no parallax & $2.86_{+0.47}^{-0.81}$ & $1.35_{+0.79}^{-0.67}$ & $0.51_{+0.38}^{-0.29}$ & $6.73_{+5.12}^{-3.02}$ \\
    ZTF19aaprbng & no parallax & $2.83_{+0.49}^{-0.80}$ & $1.26_{+0.70}^{-0.57}$ & $0.67_{+0.42}^{-0.34}$ & $4.18_{+4.10}^{-2.00}$ \\
    ZTF22aalftxa & no parallax & $2.99_{+0.38}^{-0.71}$ & $1.48_{+0.73}^{-0.65}$ & $0.59_{+0.40}^{-0.31}$ & $4.68_{+3.27}^{-2.04}$ \\
    ZTF20aauodap & no parallax & $2.95_{+0.40}^{-0.74}$ & $1.44_{+0.74}^{-0.66}$ & $0.58_{+0.41}^{-0.31}$ & $5.08_{+3.70}^{-2.20}$ \\
    ZTF20aawanug & no parallax & $2.87_{+0.47}^{-0.80}$ & $1.39_{+0.87}^{-0.74}$ & $0.40_{+0.36}^{-0.24}$ & $7.80_{+7.08}^{-3.52}$ \\
    ZTF20abgrpjg & no parallax & $2.92_{+0.43}^{-0.75}$ & $1.43_{+0.79}^{-0.69}$ & $0.50_{+0.39}^{-0.29}$ & $6.39_{+4.62}^{-2.81}$ \\
    ZTF21abmnkml & no parallax & $2.99_{+0.38}^{-0.70}$ & $1.47_{+0.70}^{-0.63}$ & $0.66_{+0.41}^{-0.34}$ & $3.73_{+2.88}^{-1.67}$ \\
    \vdots & \vdots & \vdots & \vdots & \vdots & \vdots \\
    \enddata
\end{deluxetable*}

\bibliography{Zhai}

\end{document}